\documentclass{JINST}
\usepackage{epsfig}
\usepackage{subfigure}
\usepackage{here}
\usepackage{graphicx}

\makeatletter
\renewcommand{\@thesubfigure}{\Large(\alph{subfigure})}
\renewcommand{\p@subfigure}{figure\space}
\renewcommand{\p@figure}{figure\space}
\makeatother %

\title{Development of high-gain gaseous photomultipliers for the visible spectral range}

\author{A. V. Lyashenko$^a$\thanks{Corresponding author.}, A. Breskin$^a$, R. Chechik$^a$, J. F. C. A. Veloso$^b$, J. M. F. Dos Santos$^c$, and F. D. Amaro$^c$\\
\llap{$^a$} Department of Particle Physics, Weizmann Institute of Science,\\
  76100, Rehovot, Israel\\
\llap{$^b$}University of Aveiro,\\
  3810-193 Aveiro, Portugal\\
\llap{$^c$}University of Coimbra,\\
  3004-516 Coimbra,  Portugal\\
E-mail: \email{alexey.lyashenko@weizmann.ac.il}}

\abstract{We summarize the development of visible-sensitive gaseous photomultipliers, combining a semitransparent bi-alkali photocathode with a state-of-the-art cascaded electron multiplier. The latter has high photoelectron collection efficiency and a record ion blocking capability. We describe in details the system and methods of photocathode production and characterization, their coupling with the electron multiplier and the gaseous-photomultiplier operation and characterization in a continuous mode.
We present results on the properties of laboratory-produced K$_2$CsSb, Cs$_3$Sb and Na$_2$KSb photocathodes and report on their stability and QE in gas; K$_2$CsSb photocathodes yielded QE values in Ar/CH$_4$(95/5) above 30\% at wavelengths of 360-400 nm. The novel gaseous photomultiplier yielded stable operation at gains of 10$^5$, in continuous operation mode, in 700 Torr of this gas; its sensitivity to single photons was demonstrated. Other properties are described.
The successful detection of visible light with this gas-photomultiplier  pave ways towards further development of large-area sealed imaging detectors, of flat geometry, insensitive to magnetic fields, which might have significant impact on light detection in numerous fields.
}

\keywords{ Gaseous detectors; Photon detectors for UV, visible and IR photons (gas) (gas-photocathodes, solid-photocathodes); Micropattern gaseous detectors (MSGC, GEM, THGEM, RETHGEM, MICROMEGAS, InGrid, etc)}

\begin{document}

\section{Introduction.}
Large-area position-sensitive fast photon detectors, with single-photon sensitivity in the visible spectral range, would lead to a significant progress in measuring light in numerous applications; major fields are particle-physics and astrophysics. Most commonly-used devices are vacuum photomultipliers (PMTs), with rather limited detection area and with bulky geometry due to mechanical constrains on the glass vacuum envelop. A possible alternative could be the use of gas-filled photomultipliers (GPMs) operating at atmospheric pressure; such devices have been successfully applied for UV-photon imaging in Cherenkov detectors \cite{chechik:08}. Among the expected virtues of visible-sensitive GPMs (VSGPMs) are large detection area, compact flat geometry, sub-mm 2D spatial resolution, fast (ns-scale) response and immunity to magnetic fields. Efforts to realize this approach have been ongoing for almost 2 decades, gradually overcoming basic and technological obstacles related to the chemical and physical fragility of the photocathodes and their vulnerability to secondary avalanche-induced processes \cite{edmends:88, peskov:99, shefer:02, balcerzyk:03}.

VSGPMs comprise a photocathode (PC) coupled to a gaseous electron multiplier; the latter is preferably a cascade of hole-multipliers \cite{chechik:08}, e.g. Gas Electron Multiplier (GEM) \cite{sauli:97} and Micro-Hole \& Strips Plate (MHSP) \cite{veloso:00} elements. Compared to "open-geometry" multipliers, e.g. wire, parallel-mesh and other multiplying structures, cascaded hole-multipliers are opaque to gas-avalanche-photons, thus preventing secondary photoelectron emission from the photocathode.  The most difficult issue, that was preventing high-gain operation of VSGPMs is the flow of avalanche ions from the amplification region back to the PC \cite{mormann:03, moermann:thes}. Impinging on the PC surface, these ions have a non negligible probability of releasing secondary electrons, which in turn initiate secondary avalanches - known as ion feedback; the latter limits the detector's gain by diverging into discharges and induces damage to the PC \cite{moermann:thes}.

Extensive studies have been carried out aiming at developing ways to reduce the ion back-flow with cascaded electron multipliers; the main results are reported elsewhere \cite{bondar:03, sauli:06, maia:04, lyashenko:06, lyashenko:07}. Recent studies showed that, at appropriate operation conditions in a cascaded electron multiplier comprising different hole-multipliers with patterned surfaces, the fraction of ions back-flowing to the photocathode (Ion-Backflow Fraction - IBF) could be reduced down to $3\cdot10^{-4}$ with a cascade of a Flipped-Reversed-Micro-Hole \& Strip Plate (F-R-MHSP), followed by a GEM and a MHSP \cite{lyashenko:07}; namely, about 30 residual ions would hit the photocathode for a total of 10$^5$ initial avalanche ions  initiated by a single photoelectron. This constitutes a record in blocking back flowing ions in gaseous detectors, applicable to other gaseous tracking devices, with particular importance for TPCs; as shown below, this low IBF value fulfills the requirement for stable operation of VSGPMs \cite{lyashenko:09}.

In this work we demonstrate, for the first time, the possibility of reaching gains of $\sim10^5$ in \textit{continuous}-mode operation in a VSGPM. We describe in details all the stages in the development, production and characterization of such devices. These include understanding of the physical processes of charge transport and secondary electron emission, reduction of the ion back-flow, photocathode production and characterization, photocathode operation in a gas environment, operation in pulsed-gated mode and, finally, continuous-mode operation at high gain of a cascaded gaseous multiplier with a visible-sensitive bi-alkali PC.

\section{Experimental setup and methods}
We describe below the methods and apparatus for photocathode production and characterization, detector assembly and its operation in an unsealed gas-photomultiplier prototype.

\subsection{General overview}
A dedicated, 3-chamber ultra-high vacuum (UHV) transfer system was designed and built for the production and characterization of alkali-antimonide photocathodes and their transfer to electron multipliers. The system permits to seal photocathodes to detector packages, with hot indium-bismuth alloy, as described in \cite{moermann:thes, balcerzyk:03}. The research results presented here were obtained in a non-sealed detector setup.

\begin{figure}[!ht]%
\begin{center}%
\subfiguretopcaptrue
\subfigure[][] % caption for subfigure a
{
    \label{fig:UHV:scheme}
    \includegraphics[width=14cm]{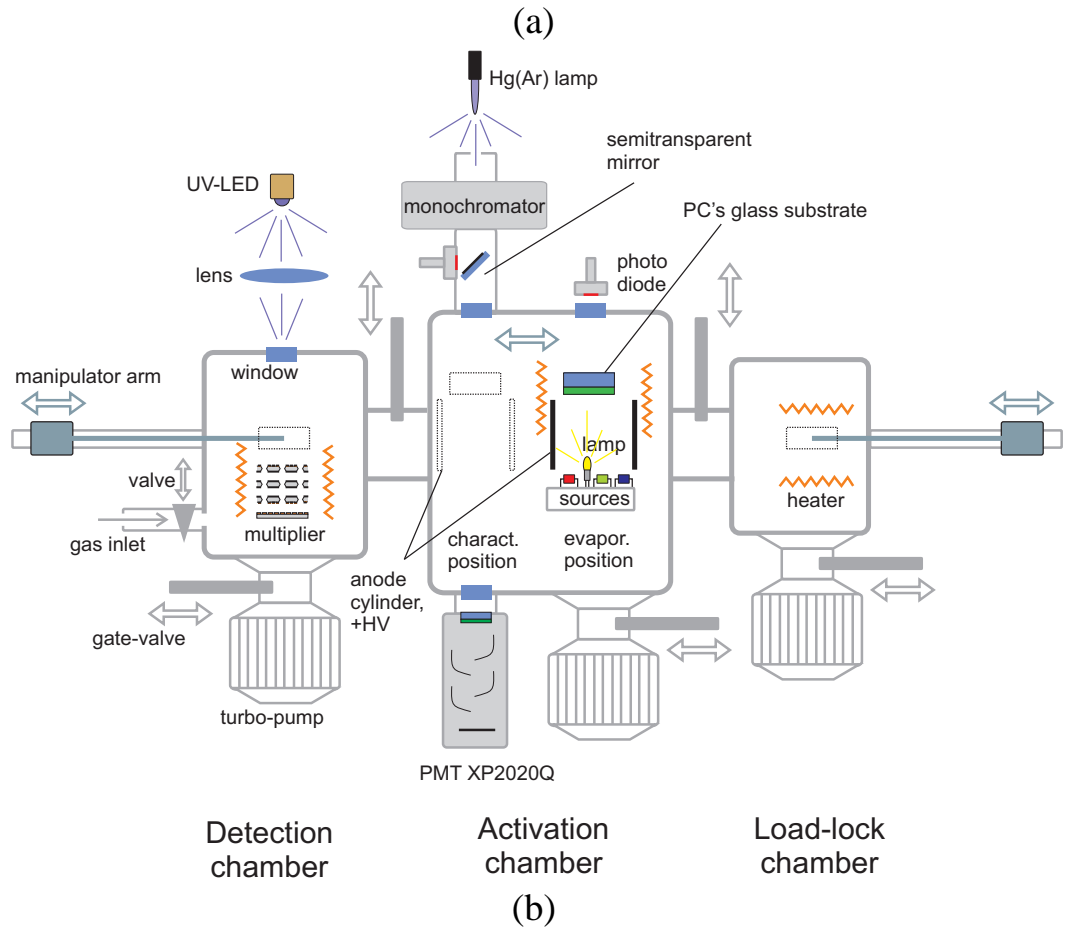}
} \hspace{0.2cm}
\subfigure[][] % caption for subfigure b
{
    \label{fig:UHV:photo}
    \includegraphics[width=10cm]{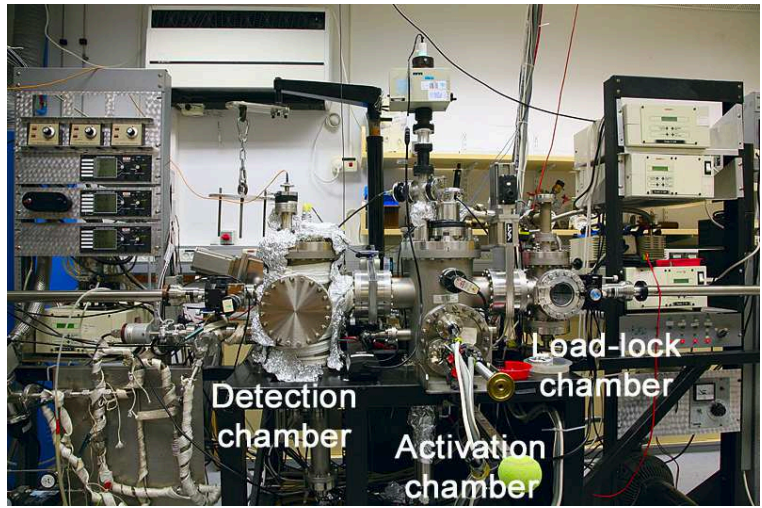}
} \caption{a) The 3-chamber UHV setup for gaseous-photomultiplier studies. From right to left: a load-lock chamber for the photocathode-substrate baking; an \textit{activation chamber} for the photocathode preparation and characterization; the \textit{detection chamber} for the characterization of a gaseous multiplier coupled to the photocathode (also including a detector sealing facility). b) A photograph of the UHV setup. }
\label{fig:UHV:setup}
\end{center}
\end{figure}

A schematic illustration and a photograph of this setup are shown in \ref{fig:UHV:setup}. The three ultra-high vacuum chambers are separated by gate valves and connected with linear magnetic manipulators; each chamber is evacuated by a separate turbo-molecular pump and baked out for $>$48 hours prior to operation. The load-lock chamber is used for introducing and baking out photocathode-substrates prior to their transfer into the second, \textit{activation chamber}. Here, the alkali-metals are evaporated onto the glass substrate to form a semiconductor photocathode. In the same chamber, the QE of the photocathode is measured in-situ. In the third \textit{detection chamber}, the electron multiplier is introduced, tested and baked before it is sealed, in gas, to a photocathode. In this work, the \textit{detection chamber} was used for the characterization of electron multipliers in combination with alkali-antimonide photocathodes - without sealing the two components. The individual elements of the system and the respective methodologies are described in detail in the following sections

\subsection{The load-lock chamber}

Designed for introducing and baking out of photocathode glass-substrates prior to their transfer into the \textit{activation chamber}, the load-lock chamber is baked out with internal quartz lamps up to 220$^0$C and is evacuated by a turbo-molecular pump, backed by a diaphragm pump. Typically, a pressure of $\sim5\cdot10^{-8}$ Torr is reached after a bake-out at 200$^{0}$C for 48 hours. The temperature of the chamber is monitored by a thermocouple, which is positioned in the vicinity of the photocathode holder.

\subsection{The activation chamber}
Photocathode production and characterization takes place inside the \textit{activation chamber}. It is baked out internally by quartz lamps before photocathode production; it is evacuated by a turbo-molecular pump, backed by a scroll pump. A base pressure of $3\cdot10^{-10}$ Torr is typically reached after a bake-out at 250$^{0}$C for 48 hours. Additional quartz lamps are located within the chamber close to the photocathode-substrate holder, allowing for local heating during the photocathode production process. The temperature in the chamber is monitored with thermocouples placed close to the substrate holder.

\begin{figure}[!ht]%
\begin{center}%
\subfiguretopcaptrue
\subfigure[][] % caption for subfigure a
{
    \label{fig:anode:comp}
    \includegraphics[width=7cm]{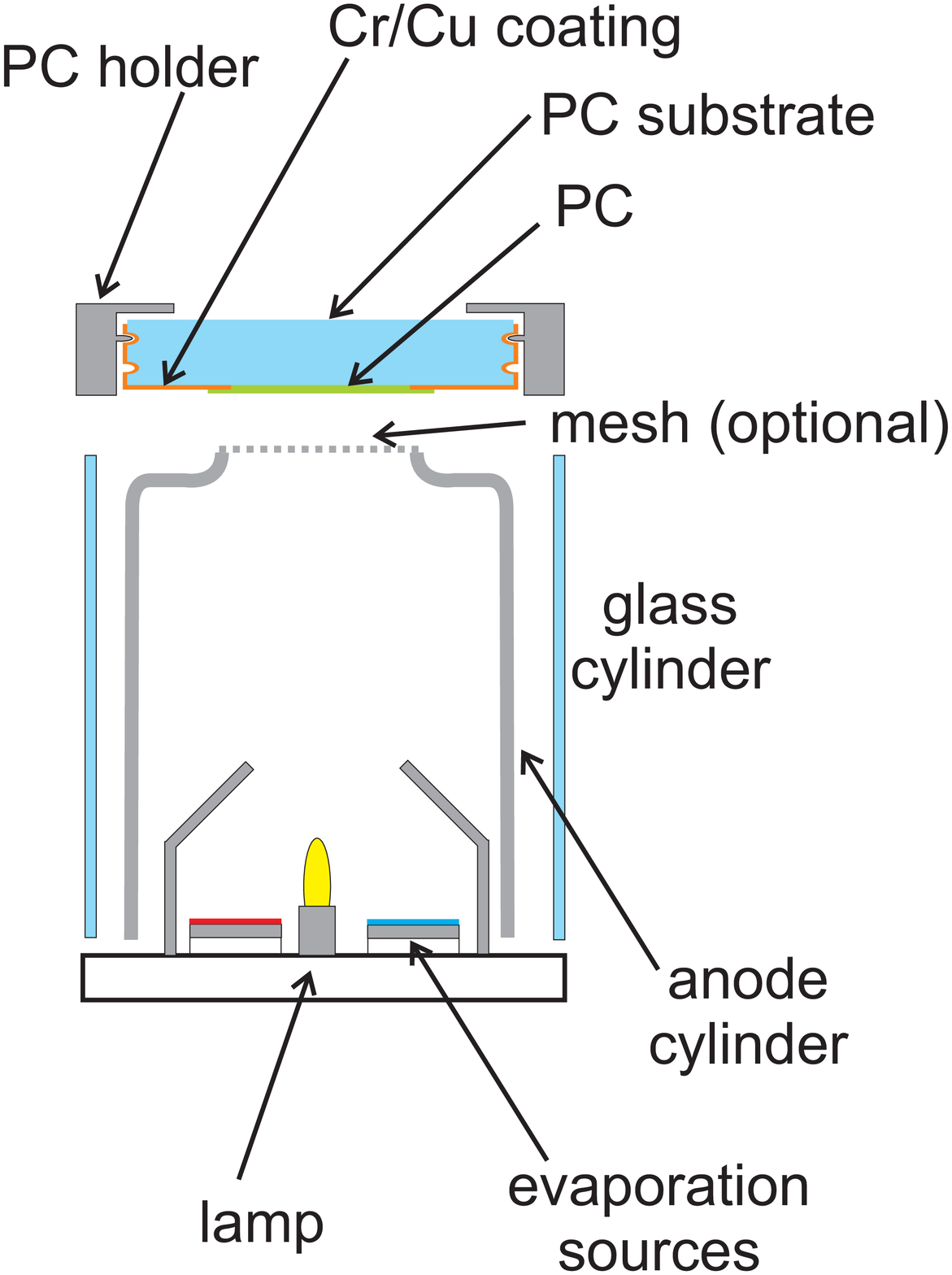}
} \hspace{0.2cm}
\subfigure[][] % caption for subfigure b
{
    \label{fig:PC:subs}
    \includegraphics[width=7cm]{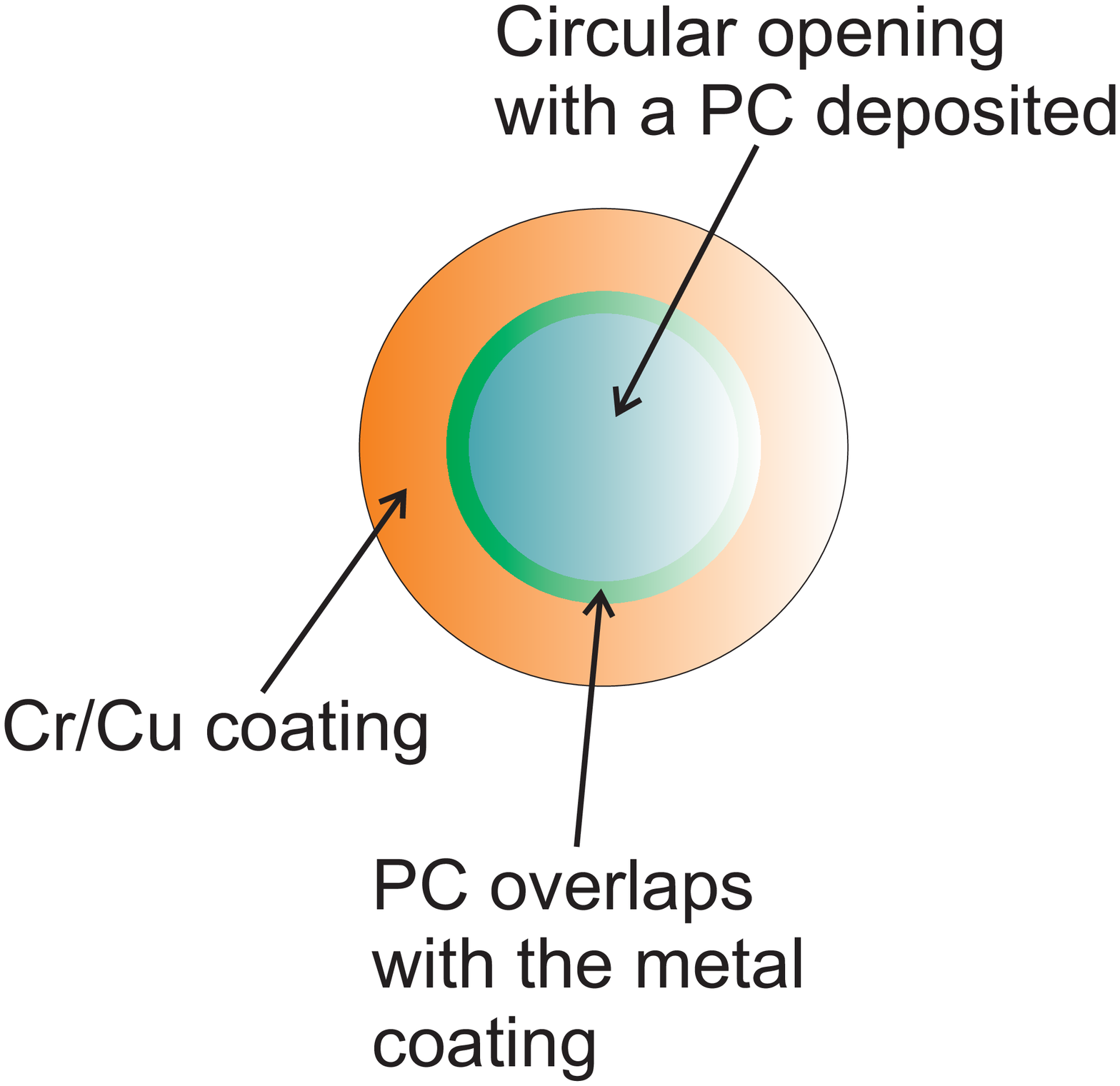}
} \caption{a) Schematic side view of the photocathode (PC) substrate placed in the evaporation position within the activation chamber. The PC substrate is shown with the Cr/Cu metal coating to provide electrical contact between the PC and the substrate holder; it has grooves for a good grip of the PC substrate by the substrate-holder. b) Sketch of the top view of the PC substrate showing the metal contact and the evaporated PC surface.}
\label{fig:PC:anode}
\end{center}
\end{figure}

The \textit{characterization position} (\ref{fig:UHV:scheme} and \ref{fig:anode:comp}) consists of three separate evaporation stations placed on a moving arm; each of them permits the production of up to three bi-alkali photocathodes. Each station contains Sb, K and Cs (or Na) evaporation sources and a small incandescent lamp used for light transmission measurement during the photocathode processing. Several antimony shots each of 1-2 mm in diameter are placed in a small Ta evaporation boat; they are pre-melted in high vacuum before installation in the \textit{activation chamber}. K, Cs and Na evaporation sources are provided by the manufacturer (SAES Getters S. p. A.) in form of small dispensers, three of each are interconnected in series by spot-welding and are placed in the respective source holders. All sources are out-gassed during the bake-out process by resistive heating. A shutter above the evaporation sources allows stopping the evaporation process during photocathode processing. A sapphire window in the \textit{activation chamber} above the evaporation position allows the illumination of the photocathode during processing (\ref{fig:UHV:scheme}).

The \textit{characterization position} (\ref{fig:UHV:scheme}) allows the in-situ measurement of the absolute QE of the photocathode: a calibrated photomultiplier (XP2020Q by Photonis Inc.) operated in photodiode mode (gain=1) is placed on the sapphire window below the \textit{characterization position}, a monochromator is placed on the top sapphire window. A Hg(Ar) lamp is used as light source for the monochromator; it exhibits narrow spectral lines in the sensitivity range of bi-alkali photocathodes (254.6, 312.5, 365.0, 404.6, 435.8, 546.0 nm). A fraction of the light on the path from the monochromator to the sapphire window is reflected by a semi-transparent mirror onto a photodiode - monitoring the lamp's light intensity.

Within the \textit{activation chamber}, the substrate and its holder are placed on top of a glass cylinder containing a cylindrical stainless-steel anode. The PC substrate in the evaporation position is shown in \ref{fig:anode:comp}. In some cases, a fine stainless steel mesh was placed on top of the anode cylinder, e.g. for creating a high homogenous electric field to reach charge multiplication for photocathode-aging studies (section \ref{sec:PC:stab}). The substrate together with the anode can be displaced between the \textit{evaporation} and \textit{characterization positions} of the \textit{activation chamber} by means of a manipulator arm. The glass cylinder confines the evaporation vapors, preventing contamination of the chamber and protecting the photocathode form eventual pollutants out-gassing from the vacuum-chamber walls. The PC substrate-holder and the anode-cylinder are electrically connected to the outside of the chamber for current measurements or for applying a high voltage.

\subsection{The detection chamber}
\begin{figure}[t]%
\begin{center}%
\subfiguretopcaptrue
\subfigure[][] % caption for subfigure a
{
    \label{fig:FRM:G:M:setup}
    \includegraphics[width=7cm]{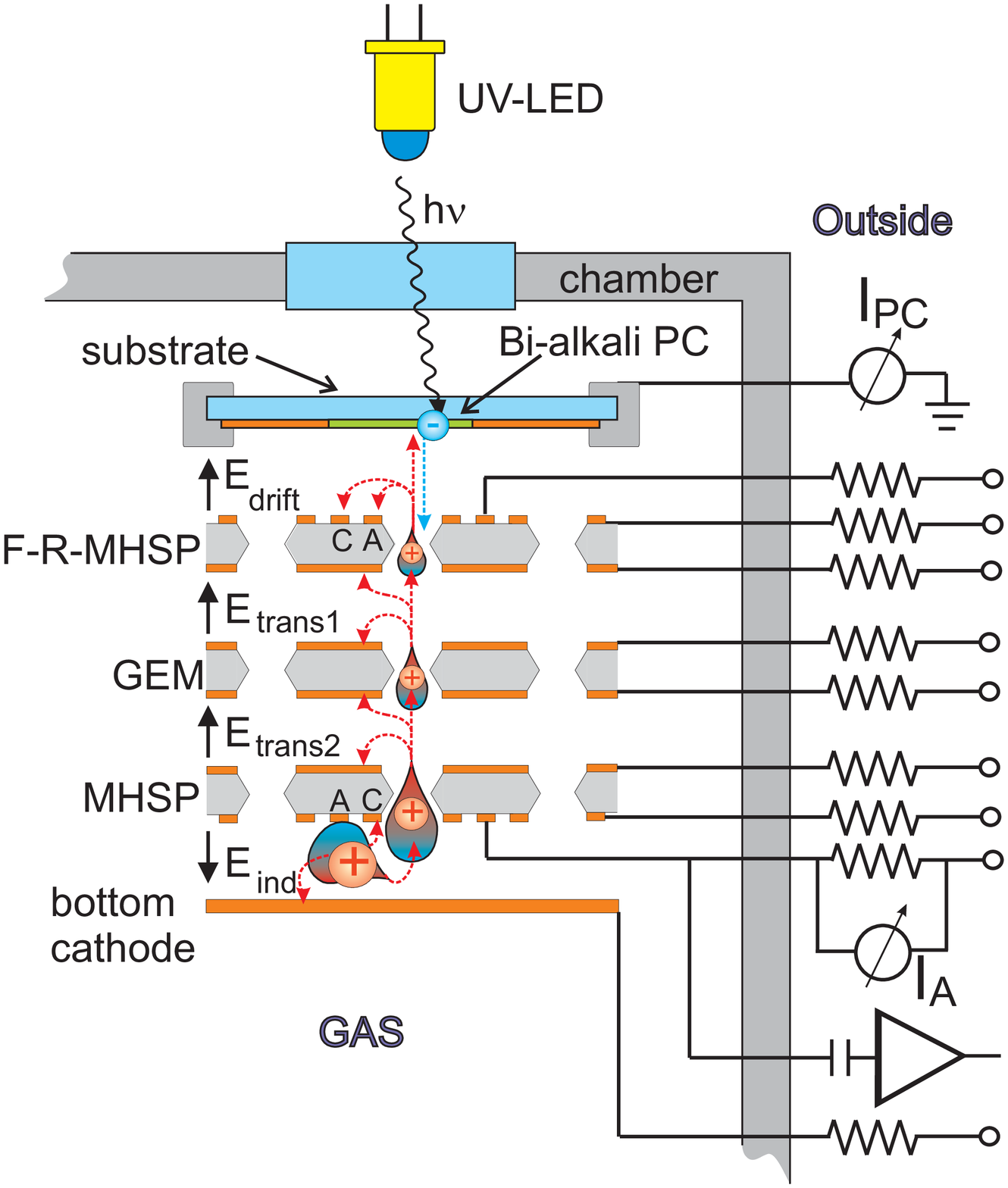}
} \hspace{0cm}
\subfigure [][] % caption for subfigure b
{
    \label{fig:det:CH3}
    \includegraphics[width=5cm]{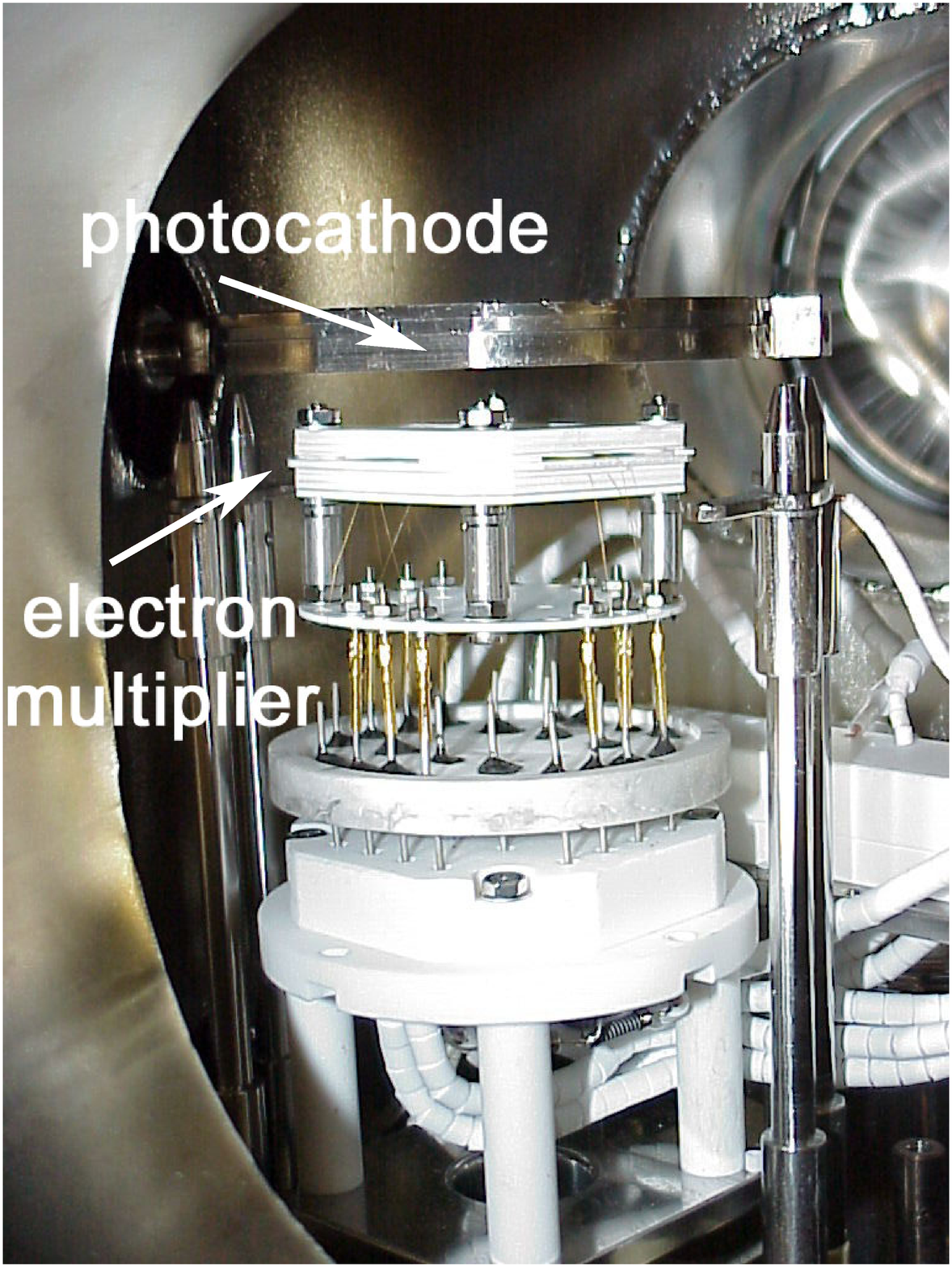}
}
\caption{a) Schematic view of a F-R-MHSP/GEM/MHSP gaseous photomultiplier assembly in the detection chamber, with a semitransparent photocathode. Photoelectrons are extracted from the photocathode into gas; they are efficiently focused into the holes of F-R-MHSP and multiplied; avalanche electrons are transferred into the GEM holes and then to the MHSP for the final multiplication. The avalanche ions (their possible paths are depicted by dotted arrows), in turn, drift back following the same electric field lines; some of them are captured by various electrodes and only a minor fraction reaches the PC. The GPM's gain and ion back-flow were established in \textit{continuous} operation mode by recording currents at the MHSP anode and at the PC; in the counting operation mode, the pulses were recorded at the MHSP anode, capacitively decoupled from a charge preamplifier. b) A photograph of the detector-holder, the multiplier-assembly and the photocathode - as mounted within the \textit{detection chamber}.}
\label{fig:FGM:set:phot}
\end{center}
\end{figure}

The \textit{detection chamber} is baked out externally by heating tapes and internally by quartz lamps; it is evacuated by a turbo-molecular pump backed by a dry scroll pump. An additional titanium sublimation pump was often used to further improve the vacuum, particularly for reducing the partial pressure of water. A base pressure of $5\cdot10^{-9}$ Torr was typically reached after a 3 days of bake-out at 160$^{0}$C. A residual gas analyzer (SRS model RGA200) monitored the vacuum quality. The detector package was fixed to a dedicated holder, establishing the electrical contacts between the detector's electrodes to the electrical feedthroughs to the external electronic circuitry (\ref{fig:FRM:G:M:setup}). The photocathode-detector assembly was illuminated from top through a quartz window. The detector assembly within the detection chamber is shown in \ref{fig:det:CH3}. This detector assembly is different than the one described in \cite{balcerzyk:03}; it does not allow for detector sealing.

\subsection{The gas system}

The gas system allows filling the \textit{detection chamber} with high-purity two-component gas mixtures. Prior to gas filling, the gas manifold is evacuated for 48 hours with a turbo-molecular pump, under bake out at 200$^0$C, down to $3\cdot10^{-6}$ Torr. The gas flow and the mixture ratio are regulated by mass-flow controllers. In all experiments Ar of 99.9999\% purity and CH$_4$ of 99.9995\% purity were used, filled into the \textit{detection chamber} through a filter; the latter (GateKeeper 35K, Aeronex Inc.) is capable of purifying noble gases, N$_{2}$ and CH$_{4}$ to ppb levels at a maximum flow of 1 liter per minute.

\subsection{The electron multiplier}
The cascaded electron multiplier, mounted in the \textit{detection chamber}, comprised a F-R-MHSP followed by a GEM and a MHSP (\ref{fig:FRM:G:M:setup}); it was coupled to a visible-sensitive semitransparent PC, typically K$_2$CsSb. The multipliers' electrodes, of 28x28 mm$^{2}$ effective area, were produced at the CERN printed-circuit workshop, from 50 $\mu$m thick Kapton foil with 5 $\mu$m Au-coated copper cladding on both faces; their structures are provided in \cite{lyashenko:07}. All the components of the multiplier were UHV-compatible, including MHSP and GEM electrodes; the latter are known to be compatible with bi-alkali PCs, at least over a few months \cite{balcerzyk:03}. The detector elements were held in place and separated from each other by 1 mm thick alumina-ceramic frames with a central opening of 20x20 mm$^{2}$, defining the active area of the electron multiplier. The assembled multiplier was mounted on a holder inside the \textit{detection chamber} as shown in \ref{fig:det:CH3}.

Following the introduction of the multiplier into the \textit{detection chamber}, the entire system was baked at 160$^0$C for 5 days in high vacuum; the temperature was limited by the multipliers' Kapton substrate. After the PC deposition and characterization in the \textit{activation chamber} in vacuum, both the \textit{activation} and \textit{detection chambers} were filled, through the purifier, with Ar/CH$_4$ (95/5) gas mixture to a pressure of 700 Torr. The PC substrate was transferred from the \textit{activation chamber} and placed in the \textit{detection chamber} at 8 mm above the multiplier's top element.

The detector and the scheme of electrical connections are shown in \ref{fig:FRM:G:M:setup}. Three adjustable voltages were supplied to the MHSP and F-R-MHSP electrodes and two to the GEM electrode, all tuned to provide the multiplication fields inside the holes and the appropriate voltage drop between the strips. The electrodes were connected independently to HV power supplies (type CAEN N471A) through 40 M$\Omega$ resistors (\ref{fig:FRM:G:M:setup}); the PC was kept at ground potential.

The GPM investigations in the \textit{detection chamber} were carried out with a UV-LED light source (NSHU590A, Nishia Corp.), whose narrow spectral emission around 375 nm coincides with the sensitivity peak of the K$_2$CsSb photocathode. The LED's light was transmitted with an optical fiber and focused onto the PC of the GPM by means of a small lens through a quartz window. The GPM could be operated either in \textit{continuous} illumination mode or in \textit{pulsed} illumination mode depending on the powering scheme of the UV-LED.

In the \textit{continuous} illumination mode, a forward bias was applied to the UV-LED. The current after multiplication was recorded on the biased MHSP anode as a voltage-drop across a 40 M$\Omega$ resistor, with a Fluke 175 voltmeter having 10 M$\Omega$ internal impedance (\ref{fig:FRM:G:M:setup}). The combined resistance was 8 M$\Omega$, from which the anode current was calculated. The avalanche-induced currents were always kept well below 100 nA by attenuating the UV-LED photon flux, to avoid charging-up effects. The currents on grounded electrodes were recorded with a Keithley 485 picoamperemeter. The multiplication (gain) curves of the detectors were deduced from the ratio of the anode current ($I_A$) to the initial photocurrent emitted, without multiplication, from the PC ($I_{PC}$).

The \textit{pulsed} illumination mode was realized by applying short voltage pulses to the LED; the emitted light intensity could be conveniently controlled by adjusting the height and width of the voltage pulses and/or adding light absorbers, down to the single-photon level. The UV-LED was powered by a pulse generator (Hewlett Packard 8012B) with square pulses having typically an amplitude of 6 V, a width of 2 $\mu$s, a period of 1 ms and rise and fall times of about 10 ns. Capacitively decoupled from the anode high voltage, the charge signal was recorded by a charge-sensitive preamplifier (typically ORTEC 124) followed by a pulse-shaping linear amplifier (ORTEC 571). The pulses were either observed on a digital oscilloscope or fed into a multi-channel analyzer (Amptek MCA8000), providing pulse-height spectra.

\subsection{Photocathode production and characterization}

The production of K$_2$CsSb photocathodes in laboratory conditions was successfully established in our group \cite{shefer:98, moermann:thes, balcerzyk:03} and by others \cite{braem:03}. It requires a careful choice and design of the experimental equipment and materials used within the vacuum chambers, due to the high chemical reactivity of alkali-antimonide photocathodes. UHV conditions with very low residual-moisture content are required for the successful production of stable high-QE photocathodes. The dedicated system described above allows for the production of several semi-transparent Cs$_3$Sb or K$_2$CsSb or Na$_2$KSb photocathodes per week.

The detailed procedures of Cs$_3$Sb, K$_2$CsSb and Na$_2$KSb photocathodes fabrication presented below, were optimized and continuously refined - resulting in high-QE photocathodes with good reproducibility.

\subsubsection{Photocathode substrate preparation}
The photocathode substrate is made from Kovar-glass, ${\o}$64 x 5 mm thickness. Its thermal expansion coefficient matches that of the Kovar-made detector packages, previously used for sealed GPMs \cite{balcerzyk:03}. Two grooves were machined on the window's edge, required for fixing it to the stainless-steel substrate-holder (\ref{fig:anode:comp}). A thin metal layer with 20 mm circular opening in the center was evaporated onto one of the substrate's faces and onto the substrate's edge (\ref{fig:anode:comp} and \ref{fig:PC:subs}); it provided electrical contact to the photocathode (deposited on the central non-metallized surface); in sealed GPMs, it was required for sealing the substrate to the Kovar package. The metal layers were electron-gun deposited in vacuum; they comprised a first chromium (100 nm) film, covered by a copper (200 nm) one; the metal sources had purity of 99.999\%. The photocathode partly overlapped with the metal, ensuring good electrical contact (\ref{fig:PC:subs}). Following evaporation, the windows were either immediately installed in the load-lock chamber or stored under vacuum. The photocathode substrates were mounted in their dedicated holders during the whole process of photocathode production and characterization, including their positioning above the multiplier within the detection chamber.  The holder permitted transporting the substrate among the three chambers with the magnetic manipulators.

The substrate was first baked in the load-lock chamber, under vacuum, by the internal quartz lamps at $>200^0$C for 48 hours. After cooling to room temperature, the baked substrate was transferred into the activation chamber with the magnetic manipulator at a pressure of typically $5\cdot10^{-8}$ Torr. It was first placed in the characterization position and its light transmission characteristic was evaluated (\ref{fig:QE:measure}), as required for the calculation of the photocathode's absolute QE. The substrate was illuminated by the monochromator and the light transmitted through the substrate was measured with the PMT. The photocurrent $I_{PMTtrans}(\lambda)$ was recorded by the Keithley 485 picoampermeter for the different characteristic wavelengths of the Hg (Ar)-lamp; the lamp's intensity was monitored  by the photodiode current $I_{PDtrans}(\lambda)$, used for photon-flux compensation if required.

\subsubsection{Cs$_3$Sb photocathodes}
The production process of Cs$_3$Sb PC's is rather simple; it consists mainly of the activation of a thin antimony film in cesium vapors at high temperature. The first step was the evaporation of an antimony film onto the PC substrate. Prior to this step, the substrate was heated to 170-180$^0$C by the internal quartz lamps; after reaching this temperature, the heating was switched off. A photodiode was then placed on the top sapphire window, illuminating the photocathode substrate from below with the incandescent lamp of the evaporation station (\ref{fig:UHV:scheme} and \ref{fig:anode:comp}). The initial photodiode current recorded defined the substrate's transmission ($I_0$). By applying a current through the dispenser antimony was evaporated onto the substrate; it was empirically found, that an optimal thickness of the antimony film corresponded to a reduction of the substrate's light transmission to about 75-85\% of its initial value $I_0$ measured before deposition. Once the transmission dropped to that level, the evaporation was terminated. Prior to cesium evaporation, the monitoring photodiode was replaced by a Ar(Hg)-lamp, illuminating the photocathode substrate from above; alternatively, the UV-LED, whose narrow spectral emission around 375 nm roughly coincides with the sensitivity peak of mono- and bi-alkali photocathodes, was used.  It allowed better focusing of the incident light onto the PC's active area, reducing the amount of light reflected on various inner chamber parts, impinging on the PC. During cesium evaporation, the PC substrate was connected to a picoampermeter for monitoring the photocurrent $I_{PC}$. The photoelectrons were collected at the anode cylinder located below the PC substrate, biased at 300 V (\ref{fig:UHV:scheme} and \ref{fig:anode:comp}). Sb activation with Cs vapors was carried out at a substrate temperature of 150-180$^0$C; it lasted until a maximum photocurrent value was reached; usually the photocurrent tended to drop rapidly at this point. Experience showed that if the Cs dispenser's power is stopped (with the shutter kept open) before a 10-30\% drop of the PC current, the latter will return to its maximal value. In some cases, the PC current did not reach its maximum value after Cs evaporation was stopped, indicating an overdose of Cs. Cs excess could be removed by heating up the PC substrate to 180-200$^0$C; the heating was stopped when the PC current returned to its maximal value. An alternative method for Cs$_3$Sb PC production consisted of evaporating an excessive amount of Cs, followed by photocurrent maximization by heating the substrate. After Cs evaporation, the Cs$_3$Sb PC was cooled down to about 150$^0$C and was kept at this temperature for about 20 minutes, monitoring its photocurrent. The last step permitted forming the semiconductor-PC structure. In cases where during this process the PC current dropped, it could be re-gained by some additional evaporation steps of Sb and/or Cs.

The Cs$_3$Sb PCs are characterized by rather low surface resistivity of about $3 \cdot 10^{7}$ $\Omega/\square$ \cite{engstrom:80} as compared to that of other alkali-antimonide PCs. It is an attractive feature, if large-area active surface is required; though, the dark emission current of about 0.3 fA/cm$^2$ \cite{engstrom:80} is higher compared to other alkali-antimonides.

\subsubsection{K$_2$CsSb photocathodes}

The production technology of K$_2$CsSb PC is considerably more complex then that of Cs$_3$Sb; however, due to their lower ($\sim$10-fold) thermo-emission currents ($\leq$0.02 fA/cm$^2$ \cite{engstrom:80}) and higher quantum yield ($\sim$40\% at 370-410 nm), these PCs are widely employed in photon detectors. A drawback of K$_2$CsSb is the high surface resistivity ($6 \cdot 10^{9}$ $\Omega/\square$ \cite{engstrom:80}); a transparent conductive film or a conductive grid should be deposited onto the PC substrate prior to the evaporation of large-area photocathodes.

The first step in K$_2$CsSb PC production process is similar to that of Cs$_3$Sb. A thin Sb film was evaporated onto the substrate as described above. Then the photodiode on the top quartz window (\ref{fig:UHV:scheme}) was replaced by a Ar(Hg)-lamp or a UV-LED illuminating the photocathode substrate from above. During potassium evaporation, the substrate was constantly kept at 170-200$^0$C, with quartz lamps; by applying high current (4-5 A) to the K-dispensers, potassium was evaporated onto the photocathode substrate, forming a K$_3$Sb photocathode of which the  photocurrent was monitored; following 1-3 minutes of evaporation, the photocurrent, and therefore the QE of the K$_3$Sb PC, reached a plateau and eventually started dropping; experience showed that a sufficient K amount was deposited when $I_{PC}$ dropped to $\sim$90\% of its maximum value. The formation of the K$_3$Sb PC was followed by cesium activation.

There are several methods of activating K$_3$Sb with cesium vapors. One of them consists of keeping the K$_3$Sb PC at 250-280$^0$C for 5-10 min in order to decrease the amount of potassium in the PC, leaving room for cesium atoms. It is followed by cesium evaporation at 160-180$^0$C, until $I_{PC}$ reaches a peak; the process is terminated when $I_{PC}$ drops to $\sim$90\% of its maximum value. The formation of the K$_2$CsSb compound continues for some time after the evaporation is stopped; the resulting increase in QE is reflected by a rising photocurrent $I_{PC}$, typically exceeding (by 10-20\%) the maximum value reached during the evaporation process. Repeated Cs evaporation steps (so called yo-yo treatment) usually yielded considerably higher QE values compared to a single evaporation step.

An alternative method consists of exposing the K$_3$Sb PC alternately to cesium and antimony (the yo-yo technique) at 180-220$^0$C until the maximum photocurrent is reached. A combination of both methods is also possible.
During the study, we could not trace any considerable advantage of one of the production methods; the two methods permitted fabrication of highly-efficient PCs.

Another method for fabrication of K$_2$CsSb PCs is the so-called co-evaporation process \cite{braem:03}. The K$_3$Sb PC is formed by simultaneous evaporation of K and Sb; then it is treated in Cs vapor until the maximum photocurrent is reached. Our present setup does not permit using of a co-evaporation process.

\subsubsection{Na$_2$KSb photocathodes}

The Na$_2$KSb PC is known for a number of unique properties like very low thermo-emission ($\leq10^{-19}$ A/cm$^{2}$ \cite{engstrom:80}), capability of operating at high temperatures, up to 200$^0$C, and the lowest-known surface resistance of $\sim2\cdot10^5$ $\Omega/\square$ among alkali-antimonides. Its production technique is quite similar to that of K$_2$CsSb described above. The difference is in the last stage; after the formation of K$_3$Sb one proceeds with the evaporation of sodium instead of cesium. It is carried out at 220$^0$C, carefully monitoring the PC current; once it rises up, the process of replacement of potassium atoms in K$_3$Sb with sodium ones takes place. This process is hard to control; sometimes we found it difficult to stop it at the optimal ratio of 1:2 between K and Na atoms. Therefore, Na evaporation should be stopped upon a decrease in the speed of the photocurrent rise; an excess of Sodium will result in a poor-quality PC. In the following step, one proceeds with alternating (yo-yo) additions of antimony and potassium at 160-180$^0$C, until reaching the maximum photocurrent value. The formation of the Na$_2$KSb compounds continues for some time after the evaporation is stopped; the PC should be kept for some 20-30 minutes at 150-160$^0$C.

\subsubsection{Photocathode characterization}

\begin{figure} [!ht]
  \begin{center}
    \epsfig{file=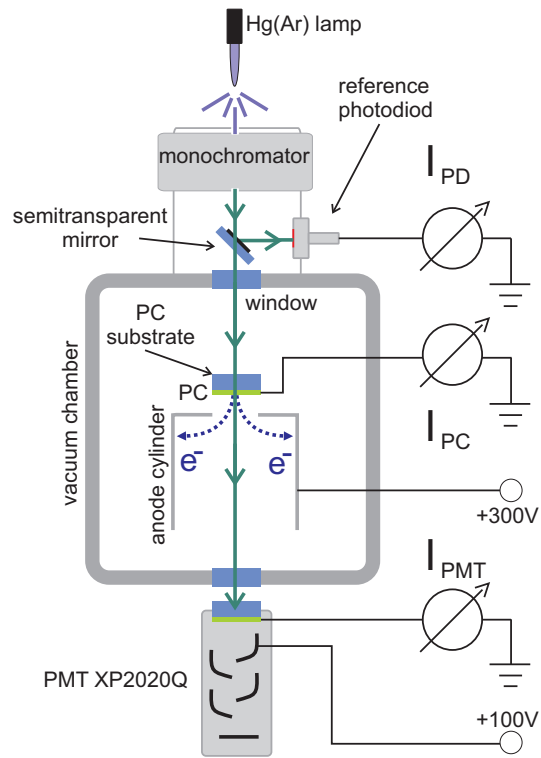, width=7cm}
    \caption{Schematic illustration of the setup for PC characterization. The PC substrate transparency is measured in the same setup prior to the PC evaporation.}
    \label{fig:QE:measure}
  \end{center}
\end{figure}

After the photocathode has cooled down to room temperature (typically 10 hours in vacuum), it is moved to the \textit{characterization position} and a positive voltage of 300 V is applied on the anode's cylinder (\ref{fig:QE:measure}). The photocathode is illuminated from above with the monochromator (\ref{fig:QE:measure}); its photocurrent, $I_{PC}(\lambda)$ and that on the reference photodiode, $I_{PD}(\lambda)$, are measured as a function of the wavelength. The absolute QE is given by the following relation:

After the photocathode has cooled down to room temperature (typically 10 hours in vacuum), it is moved to the \textit{characterization position} and a positive voltage of 300V is applied on the anode's cylinder (\ref{fig:QE:measure}). The photocathode is illuminated from above with the monochromator (\ref{fig:QE:measure}); its photocurrent, $I_{PC}(\lambda)$ and that on the reference photodiode, $I_{PD}(\lambda)$, are measured as a function of the wavelength. The absolute QE for a certain wavelength $\lambda$ is given by the following relation:
\[
    QE(\lambda)=\frac{I_{PC}(\lambda)-I_{PC}^{dark}}{I_{PMTtrans}(\lambda)-I_{PMTtrans}^{dark}} \cdot \frac{I_{PDtrans}(\lambda)-I_{PDtrans}^{dark}}{I_{PD}(\lambda)-I_{PD}^{dark}}\cdot T_W(\lambda) \cdot QE_{PMT}(\lambda)
\]

\noindent where

$QE(\lambda)$ - photocathode quantum efficiency

$I_{PC}(\lambda)$ - photocurrent measured on the photocathode

$I_{PC}^{dark}$ - dark-current measured on the photocathode with the monochromator lamp switched off

$I_{PD}(\lambda)$ - current measured on the monitoring photodiode

$I_{PD}^{dark}$ - dark-current measured on the monitoring photodiode with the monochromator lamp switched off

$I_{PMTtrans}(\lambda)$ - current on the PMT from the substrate's transmission measurement

$I_{PMTtrans}^{dark}$  - dark-current on the PMT from the substrate's transmission measurement with the monochromator lamp switched off

$I_{PDtrans}(\lambda)$ - current on the monitoring photodiode from the substrate's transmission measurement

$I_{PDtrans}^{dark}$  - dark-current on the monitoring photodiode from the substrate's transmission measurement with the monochromator's lamp switched off

$QE_{PMT}(\lambda)$ - PMT quantum efficiency as supplied by manufacturer (see table \ref{tab:QE:PMT})

$T_W(\lambda)$ - sapphire window light transmission as supplied by manufacturer (see table \ref{tab:QE:PMT})

\begin{table}
    \begin{tabular} { | l || l | l | l | l | l |}
    \hline
    $\lambda$ [nm] & 312 & 365 & 405 & 435 & 546 \\ \hline
    $QE_{PMT}$ [\%] & 28.22 & 29.38 & 28.00 & 24.74 & 6.29 \\ \hline
    $T_W$ & 0.8 & 0.8 & 0.8 & 0.8 & 0.8 \\
    \hline
    \end{tabular}
    \caption{XP2020Q PMT quantum efficiency and light transmission through the sapphire window for the characteristic wavelengths of the Hg(Ar)-lamp.}
    \label{tab:QE:PMT}
\end{table}

\section{Results}

\subsection{Alkali-antimonide photocathodes}

\begin{figure}[t]%
\begin{center}%
\subfiguretopcaptrue
\subfigure % caption for subfigure a
{
    \label{fig:QE:CsSb}
    \includegraphics[width=7cm]{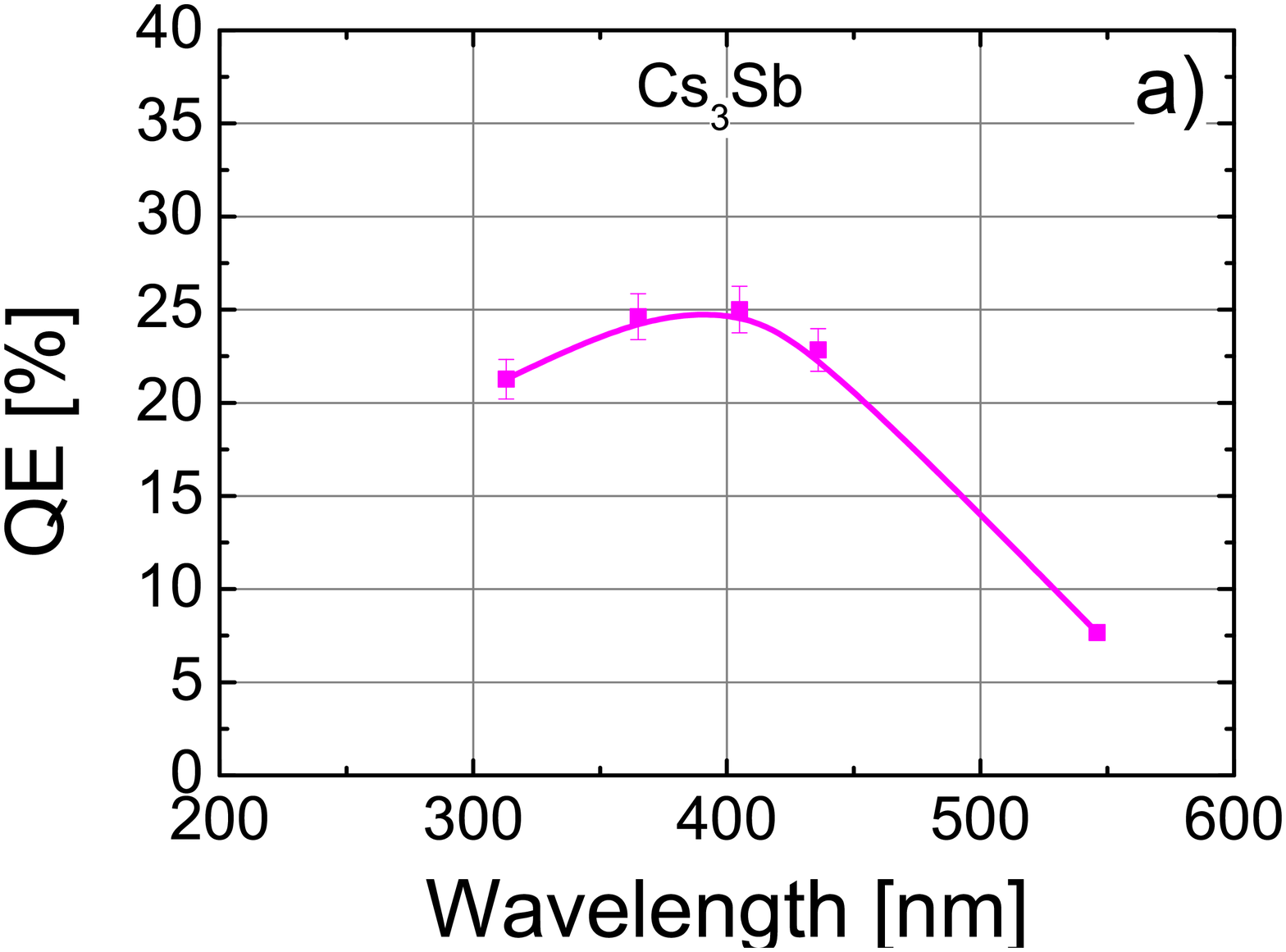}
} \hspace{0cm}
\subfigure % caption for subfigure b
{
    \label{fig:QE:KCsSb}
    \includegraphics[width=7cm]{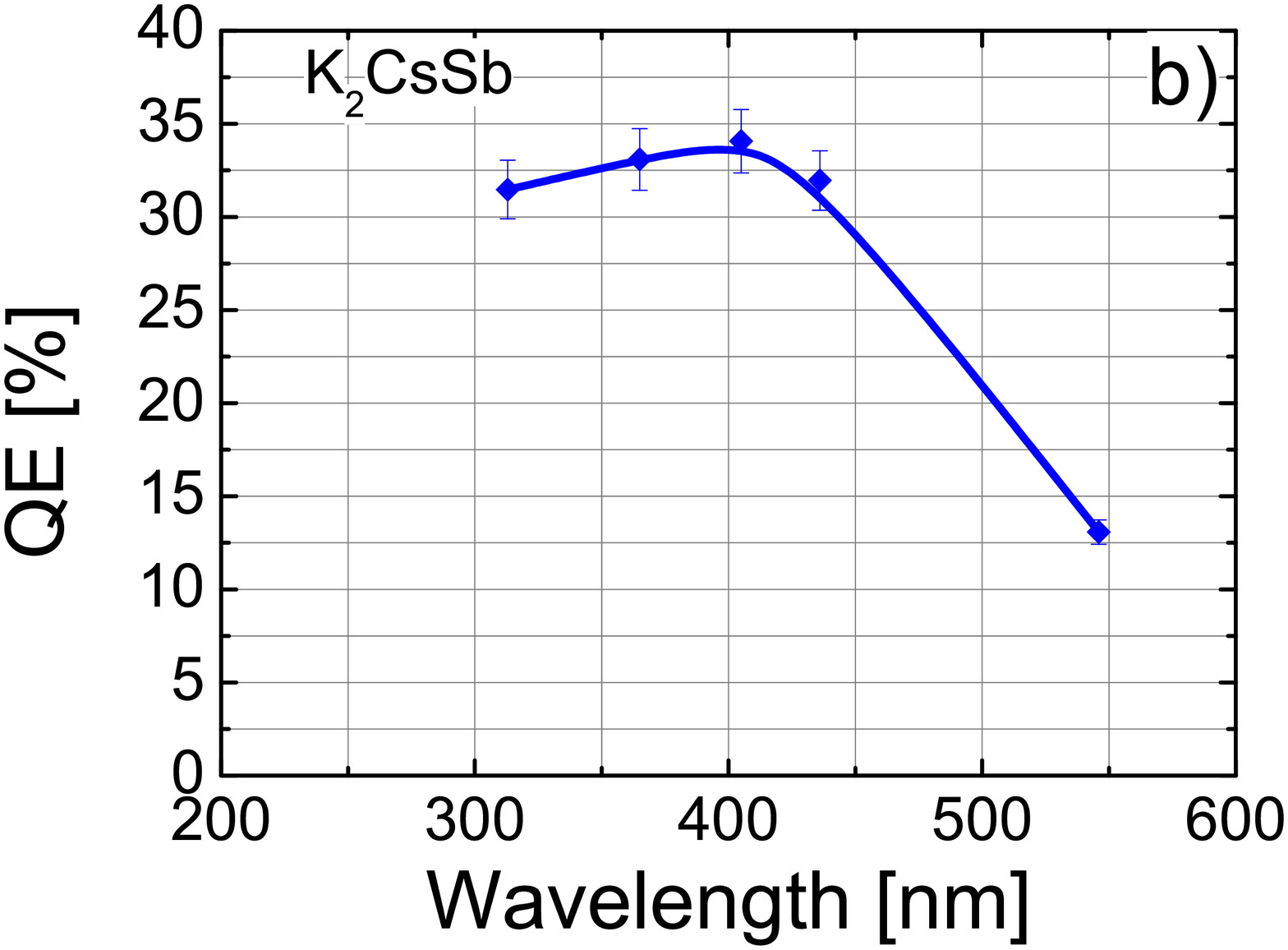}
}
\subfigure % caption for subfigure c
{
    \label{fig:QE:NaKSb}
    \includegraphics[width=7cm]{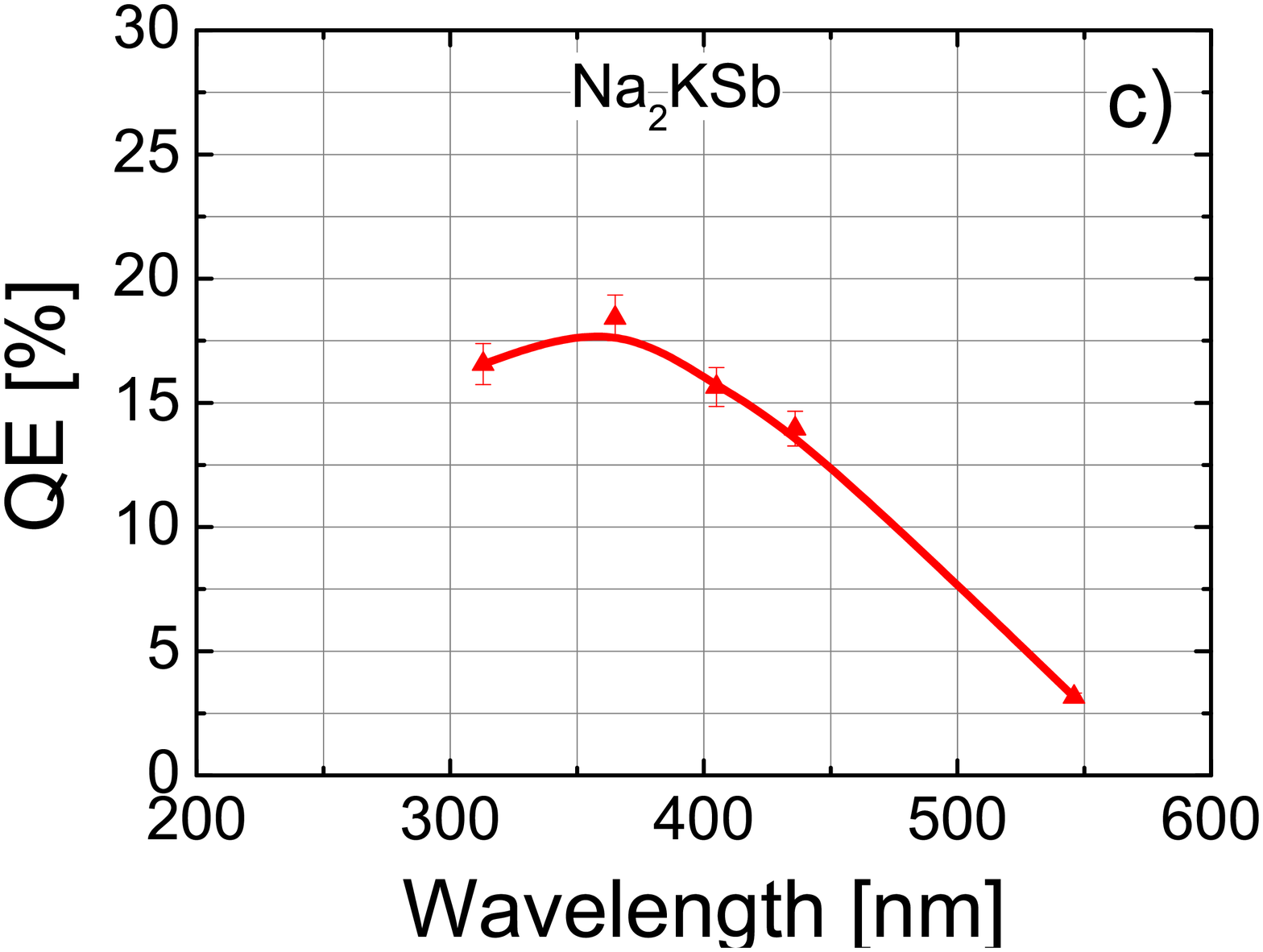}
}
\caption{Typical QE vs wavelength plots for Cs$_3$Sb a), K$_2$CsSb b) and SbK$_{2}$Na c) produced in our lab.}
\label{fig:QE}
\end{center}
\end{figure}

\begin{figure} [t]
  \begin{center}
    \epsfig{file=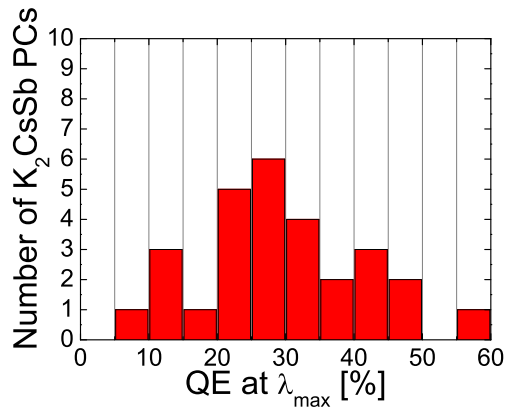, width=8cm}
    \caption{Distribution of the peak quantum efficiency values of K$_2$CsSb PCs prepared in our lab; the average QE of all the produced PCs is about 28\% (at wavelengths of 360-400 nm) }
    \label{fig:QE:dist}
  \end{center}
\end{figure}

In \ref{fig:QE} we present typical plots of QE vs wavelength for our Cs$_3$Sb, K$_2$CsSb and Na$_2$KSb PCs in the spectral range between 313 nm and 546 nm. The distribution of the peak (highest) QE-values of a large number K$_2$CsSb photocathodes prepared in our setup is shown in \ref{fig:QE:dist}; it indicates the fluctuations in the photo-emissive properties. The average QE of the produced PCs is about 28\%. As can be seen in \ref{fig:QE:dist}, out of 28 PCs produced, 23 have QE-values exceeding 20\%; the 5 PCs with QE below 20\% were mostly produced at the early stages of our studies, before mastering the production technology. The QE values increased at later stages: most of the PCs had over 20\% QE; some of them had values close to 50\% at a wavelength of 360-380 nm. Beside K$_2$CsSb PCs, we also produced 6 Na$_2$KSb PCs with peak QE varying from 15 to 25\% and 7 Cs$_3$Sb PCs with peak QE varying between 10 to 40\%.

\subsubsection{Photoemission into gas media}

\begin{figure} [!ht]
  \begin{center}
  \subfigure % no caption for subfigure a
{
    \label{fig:extr:KCsSb1}
    \includegraphics[width=7cm]{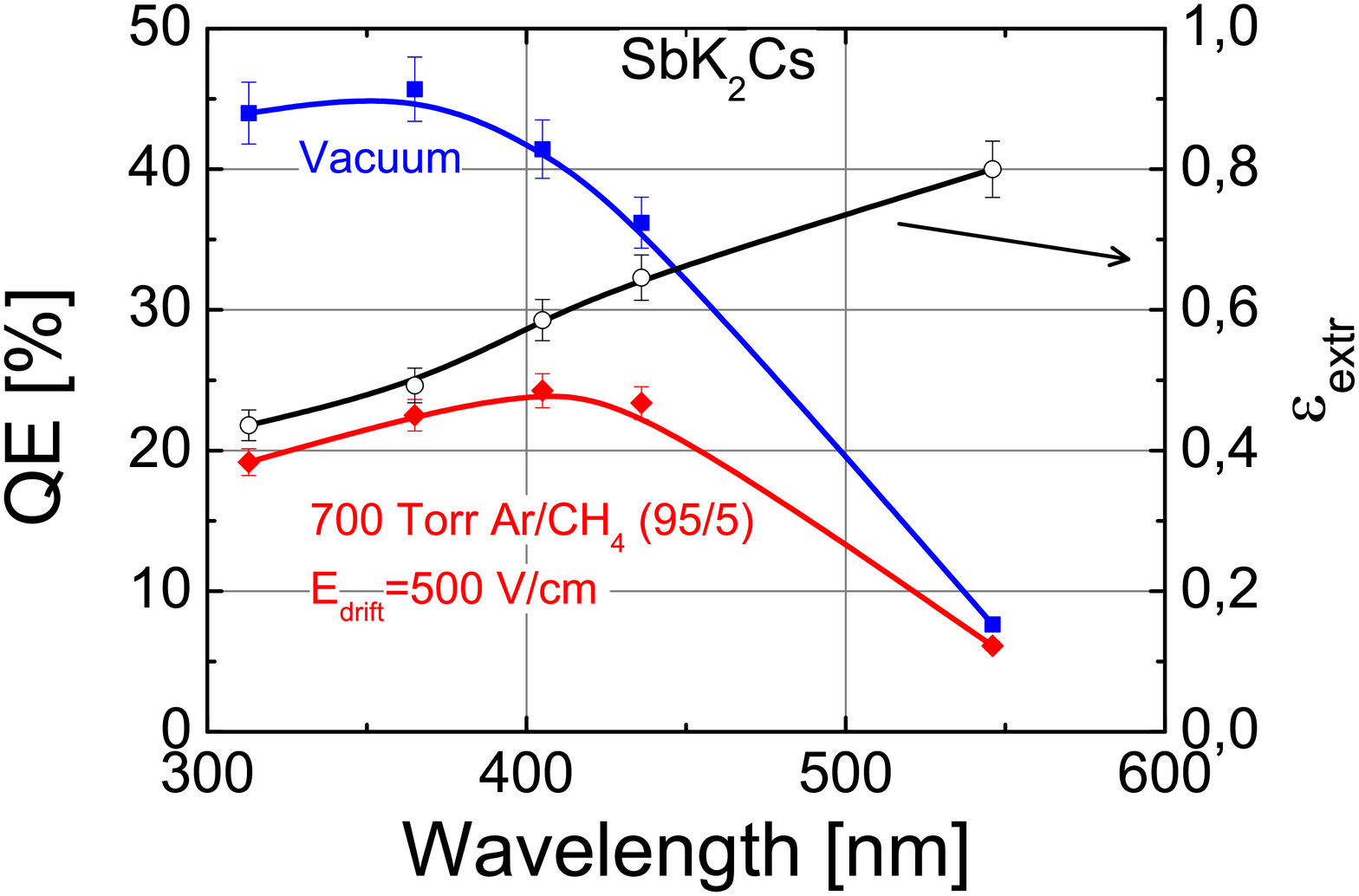}
}
\subfigure % no caption for subfigure b
{
    \label{fig:extr:KCsSb2}
    \includegraphics[width=7cm]{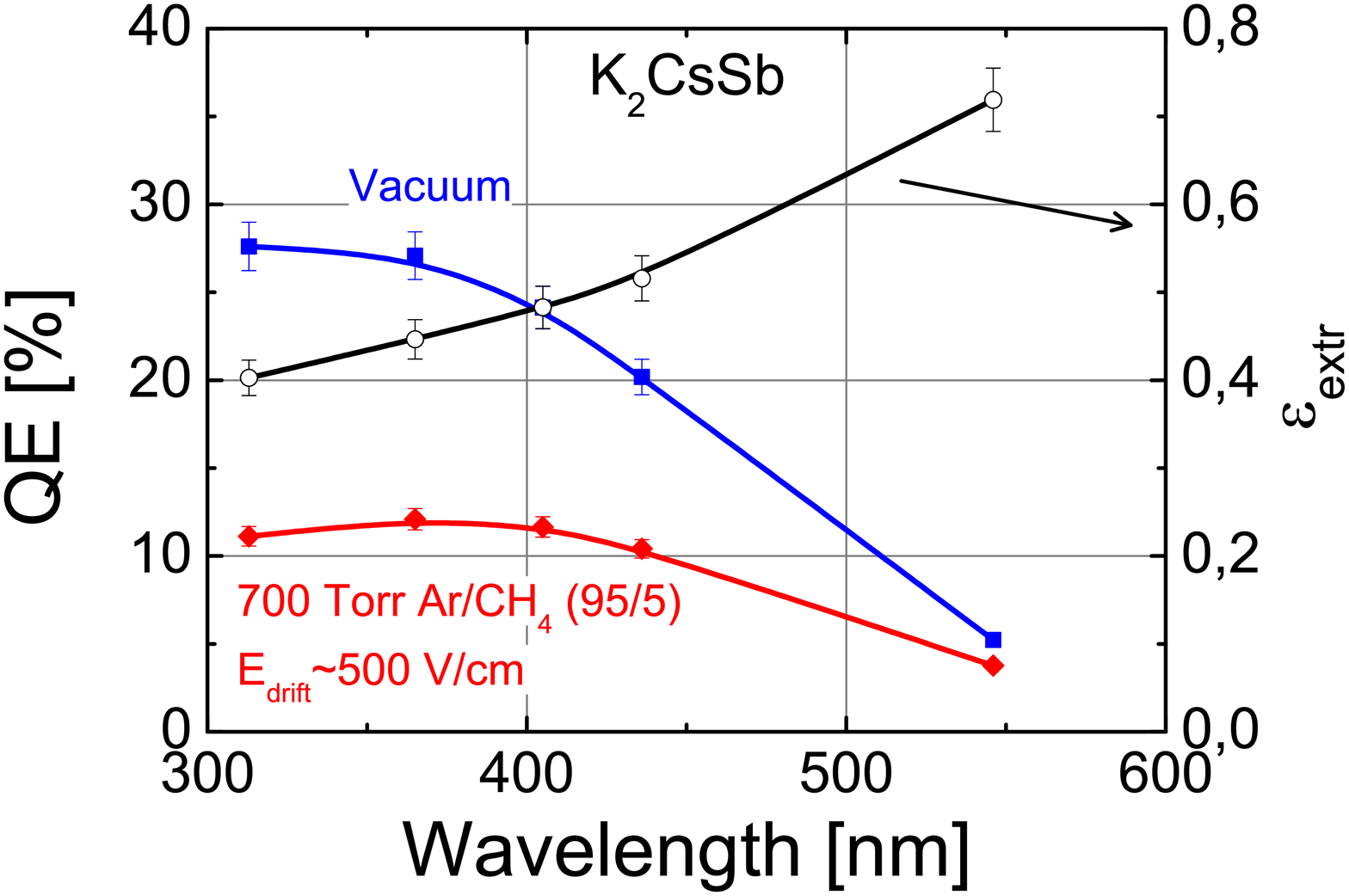}
}
    \caption{Two examples of quantum efficiency (left y-axis) vs. photon wavelength measured in 700 Torr Ar/CH$_4$ (95/5) (diamonds) and in vacuum (squares), from which the photoelectron extraction efficiency $\varepsilon_{extr}$ (circles, right y-axis) is calculated. The electric field was 500V/cm. }
    \label{fig:extr:wl}
  \end{center}
\end{figure}

The photoemission into gas differs from that in vacuum as the photoelectrons emitted from the PC are subject to scattering on gas molecules, resulting in a non-negligible fraction of them being scattered back and trapped at the PC surface; this is called the electron backscattering and it results in $QE_{gas}$ being smaller than $QE_{vacuum}$. The backscattering probability depends on the electric field at the PC surface and even more on the gas type and the kinetic-energy of the emitted photoelectrons \cite{lyashenko:09}. While photoelectron emission into gas from UV-sensitive CsI PCs was thoroughly studied both theoretically \cite{escada:07} and experimentally \cite{buzulutskov:00, dimauro:96}, similar studies of emission into gas from alkali-antimonide PCs have not been performed so far. The effect of photoelectron backscattering from K$_2$CsSb PC is illustrated in \ref{fig:extr:wl}, highlighting the difference between $QE_{vacuum}$ and $QE_{gas}$. The efficiency $\varepsilon_{extr}$ of electron extraction from the PC or the fraction of photoelectrons surmounting the backscattering was measured in the experimental setup of \ref{fig:QE:measure} with a fine stainless steel anode mesh placed on top of the metal cylinder to provide a uniform electric field in the gap between the PC and the mesh. $\varepsilon_{extr}$ was measured at an electric field of 500 V/cm, known to provide reasonable extraction efficiency of photoelectrons from CsI PC into various gases \cite{moermann:thes}; The $\varepsilon_{extr}$ values for K$_2$CsSb PCs into Ar/CH$_4$ (95:5) at 700 Torr range between $\sim$43\% at 313 nm and $\sim$79\% at 546 nm.

\subsubsection{Photocathode Stability in gas}
\label{sec:PC:stab}

The stability of alkali-antimonide PCs in gas environment was studied in the past \cite{moermann:thes, mormann:03, shefer:98, rodionov:02}. A K$_2$CsSb PC sealed in a Kovar envelop showed stable behavior at 680 Torr of pure argon during half year period \cite{mormann:03}; both Cs$_3$Sb and K$_2$CsSb PCs showed no deterioration during two-days storage in a vacuum chamber filled with pure methane at 760 Torr \cite{shefer:98}. A stable operation of a Cs$_3$Sb PC in pure Xe during 45 days was reported in \cite{rodionov:02}.

\begin{figure}[!ht]%
\begin{center}%
\subfiguretopcaptrue
\subfigure % caption for subfigure a
{
    \label{fig:stab:SbCs}
    \includegraphics[width=7cm]{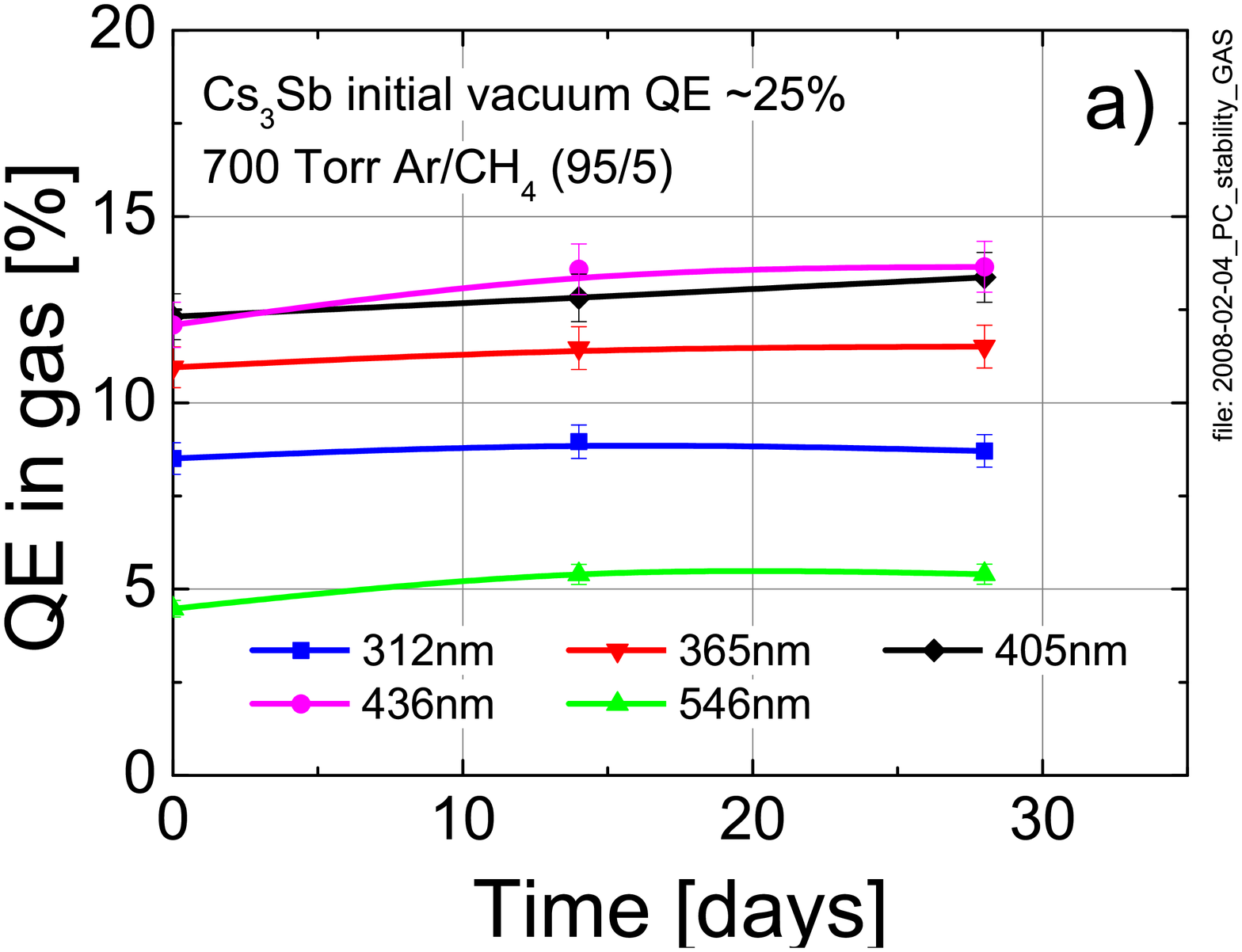}
} \hspace{0.2cm}
\subfigure % caption for subfigure b
{
    \label{fig:stab:SbKCs}
    \includegraphics[width=7cm]{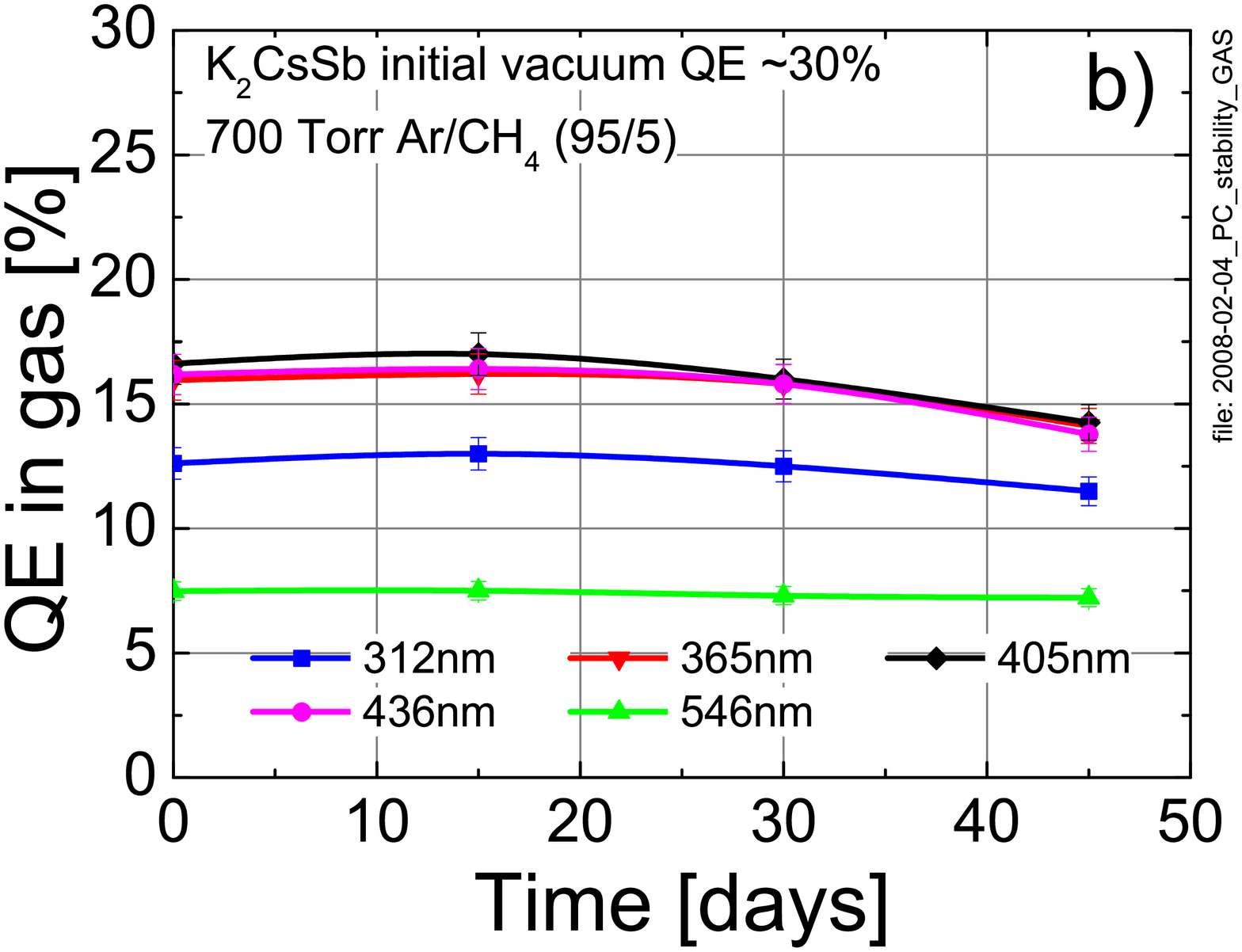}
} \caption{The time-evolution of the QE at different wavelengths of Cs$_3$Sb a) and K$_2$CsSb b) photocathodes in high purity Ar/CH$_{4}$ (95/5) at 700 Torr.}
\label{fig:stability}
\end{center}
\end{figure}

The present K$_2$CsSb and Cs$_3$Sb PCs were found to be stable in the activation chamber filled with high-purity Ar/CH$_4$ (95/5) at 700 Torr for about a month. \ref{fig:stability} shows the evolution in time of the QE in Ar/CH$_4$ (95/5) of a Cs$_3$Sb PC \ref{fig:stab:SbCs} with an initial vacuum QE-value of $\sim$25\% and of a K$_2$CsSb PC \ref{fig:stab:SbKCs} with an initial vacuum QE-value of $\sim$30\%. As shown in \ref{fig:stab:SbCs}, no appreciable change in QE was observed after 28 days for the Cs$_3$Sb PC. The K$_2$CsSb PC was stable in gas for one month; after 45 days a minor QE decay was observed which we believe is due to the impurities present in our large vacuum chamber. A sealed device is expected to have fewer impurities and thus better stability.

\subsection{High gain operation of visible-sensitive GPMs (VSGMPs)}

\subsubsection{Operation in continuous-illumination mode}

Following the success in ion back-flow reduction \cite{lyashenko:07}, a good understanding of the ion-induced secondary emission (IISEE) process \cite{lyashenko:09} and after mastering the technique for producing highly efficient alkali-antimonide PCs, we proceeded in assembly and characterization of VSGPMs in continuous operation mode. The latter comprised F-R-MHSP/GEM/MHSP cascaded multipliers installed in the \textit{detection chamber} (\ref{fig:FRM:G:M:setup}); the K$_2$CsSb evaporated in the \textit{activation chamber} was transferred with a manipulator into the \textit{detection chamber}, under gas, and placed above the multiplier. The operation principle of such VSGPM is schematically depicted in \ref{fig:FRM:G:M:setup}; photoelectrons drift into the first multiplier under a drift field $E_{drift}$ undergo avalanche multiplication in the holes, avalanche electrons are further transferred under a transfer field, $E_{trans1}$ into the GEM for a second multiplication, the avalanche electrons are then extracted by the transfer field $E_{trans2}$ and transferred to the MHSP, where an additional two-stage multiplication process occurs \cite{veloso:00}. The total avalanche charge is collected at the anode of the MHSP. The avalanche ions, in turn, drift back following the same electric field lines; most of them are captured by the various electrodes and only a small fraction reaches the PC. The multiplier was operated with full efficiency of photoelectrons collection from the PC. A more detailed explanation of the IBF measurements and on the operation of such a cascaded multiplier could be found in \cite{lyashenko:07}. The QE distribution of the K$_2$CsSb PC used in one of the experiments is shown in \ref{fig:extr:KCsSb2}.

\begin{figure} [!ht]
  \begin{center}
  \makeatletter
    \renewcommand{\p@figure}{figure\space}
  \makeatother
    \epsfig{file=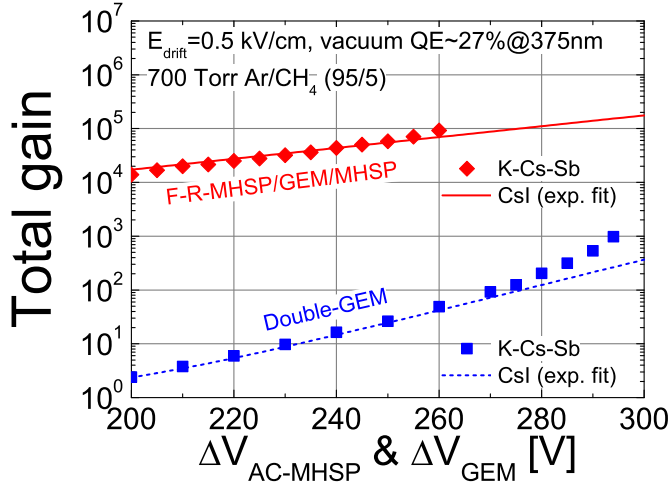, width=10cm}
    \caption{Gain curves measured in the GPM of \protect \ref{fig:FRM:G:M:setup} and, for comparison, in a GPM of 2 cascaded GEMs coupled to a semitransparent photocathode. In addition to the curves with bi-alkali photocathodes (data points only), are shown lines fitted to the data measured with CsI UV-photocathodes in the same conditions. Notice the gain "divergence" in the bialkali/double-GEM GPM already at gains $>$100. The data points for the K$_2$CsSb PC coincide with the CsI PC line, with no ion-feedback induced deviation. 700 Torr Ar/CH$_4$ (95/5); $E_{drift}$=0.5 kV/cm. QE-value refers to vacuum. The total gain in each of the two detectors was varied by adjusting either the strip voltage of the MHSP (top curve) or the GEM voltage (bottom curve); see text.}
    \label{fig:FGM:KCsSb:gain}
  \end{center}
\end{figure}

In \ref{fig:FGM:KCsSb:gain} we present gain-voltage characteristics for the cascaded GPM of \ref{fig:FRM:G:M:setup}; it was measured at the same experimental conditions with K$_2$CsSb and CsI PCs. In the latter, a substrate with CsI photocathode deposited in another system was introduced from the outside into the detection chamber.  The gain-voltage curve measured with CsI PC in \ref{fig:FGM:KCsSb:gain} shows an exponential behavior; the data points measured with the K$_2$CsSb PC coincide with the CsI PC curve. In both cases the GPM could reach gains of 10$^5$ with no divergence from exponential, indicating at full suppression of ion feed-back effects. We adjusted the inter-strip voltage on the bottom MHSP $\Delta V_{AC-MHSP}$, to vary the total gain of the GPM. The response of a GPM composed of 2 cascaded GEMs coupled to a bi-alkali photocathode is shown for comparison; notice the gain "divergence" in the bi-alkali/double-GEM GPM, occurring already at gains above 100, compared to the regular exponential behavior obtained with the cascaded GPM of \ref{fig:FRM:G:M:setup}. The voltages $\Delta V_{GEM}$ across each GEM were adjusted simultaneously to vary the total gain.

\subsubsection{Operation in pulsed-illumination mode}

\begin{figure} [!ht]
  \begin{center}
  \makeatletter
    \renewcommand{\p@figure}{figure\space}
  \makeatother
    \epsfig{file=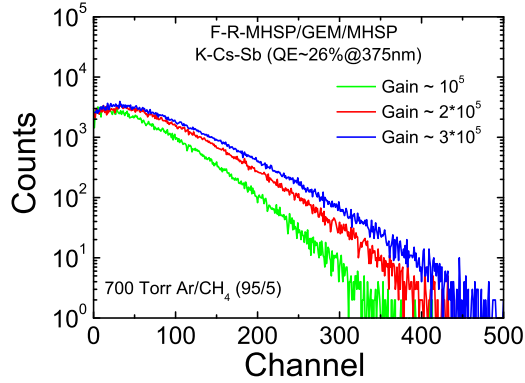, width=8cm}
    \caption{Single-electron pulse-height spectra obtained for a GPM comprised of F-R-MHSP/GEM/MHSP multiplier coupled to K$_2$CsSb PC (peak QE$\sim$27\%) with MHSP anode voltages $\Delta V_{AC}$=260, 265 and 270V, corresponding to respective gains of 10$^{5}$, $2\cdot10^5$ and $3\cdot10^5$}
    \label{fig:FGM:KCsSb:sphe:spec}
  \end{center}
\end{figure}

In pulsed-illumination mode, the detector was illuminated with short light flashes from the UV-LED. In order to demonstrate single-photon sensitivity of the GPM, the amount of photons was reduced to few photons per flash, with light absorbers and by adjustment of the pulse-width and amplitude. Pulse-height spectra of single-photon events were acquired with a F-R-MHSP/GEM/MHSP multiplier coupled to a K$_{2}$CsSb PC (peak QE$\sim$27\%) with different values of anode-to-cathode voltages $\Delta V_{AC}$ of the multiplier's MHSP. The single photon pulses were recorded at the MHSP anode capacitively decoupled from the high voltage by a charge-sensitive preamplifier followed by a pulse-shaping linear amplifier; they were then fed into a multi-channel analyzer, providing pulse-height spectra. The spectra were recorded at $\Delta V_{AC}$=260, 265 and 270V, corresponding to VSGPM gains of approximately 10$^{5}$, $2\cdot10^5$ and $3\cdot10^5$. The gains were estimated by fitting the gain-voltage curves of \ref{fig:FGM:KCsSb:gain} with an exponential function. The noise spectrum for each anode-to-cathode voltage was recorded separately at the same conditions with the UV-LED switched off. The noise spectrum was later subtracted from the experimental one for each anode-to-cathode voltage. The resulting spectra are shown in \ref{fig:FGM:KCsSb:sphe:spec}. The exponential-shape of the spectra presented in \ref{fig:FGM:KCsSb:sphe:spec} proves that the recorded pulses indeed originate from single photoelectrons. The slightly peaked distributions at the lower pulse-heights, is probably due to an over estimated (subtracted) electronic noise.

\begin{figure}[!ht]
  \begin{center}
  \makeatletter
    \renewcommand{\p@figure}{figure\space}
  \makeatother
    \epsfig{file=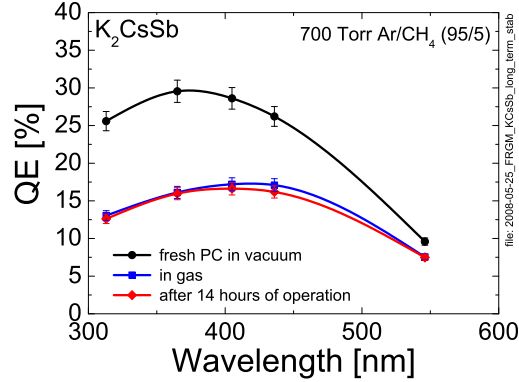, width=8cm}
    \caption{The QE of the K$_2$CsSb photocathode measured in vacuum (circles) at 700 Torr Ar/CH$_4$ (95/5), immediately after gas introduction (squares) and after 14 hours operation in detection chamber with F-R-MHSP/GEM/MHSP multiplier at a gain of 10$^5$ (diamonds).}
    \label{fig:FGM:KCsSb:QE:long:term}
  \end{center}
\end{figure}

The stability of the VSGPM was studied in the same setup. The F-R-MHSP/GEM/MHSP multiplier coupled to a K$_2$CsSb PC was illuminated with multiple-photon flashes from the UV-LED. It was illuminated for 14 hours at an average photon rate of $\sim$12kHz/mm$^2$ resulting in a total charge of $\sim$125 $\mu$C collected at the MHSP's anode at a gain of 10$^5$; no sparks were recorded during the operation time. Charge pulses were recorded at the detector's anode when starting the irradiation and after 14 hours of operation at a gain 10$^5$. The signal's amplitude decreased by $\sim$14\% of its initial value; this signal decay could be related to gain variation due to initial charging up phenomena of the multiplier's polymer substrates, currently observed in micro-pattern multipliers \cite{ahn:01, azmoun:06}, and not by the PC decay.  \ref{fig:FGM:KCsSb:QE:long:term} indeed confirms the PC stability throughout the study; it depicts the QE of the K$_2$CsSb photocathode as a function of wavelength at different stages of the experiment. The QE was measured in vacuum for the freshly produced PC, then following gas introduction (at an electric field of 500 V/cm at the PC in 700 Torr Ar/CH$_{4}$ (95/5)) and after 14 hours of operation in the detection chamber with the F-R-MHSP/GEM/MHSP multiplier at a gain of 10$^5$. The QE was practically not affected during its operation with the multiplier; the small decrease in QE observed at 436 nm is within the measurement accuracy. Details about previous PC decay studies under avalanche-ion bombardment can be found elsewhere \cite{breskin:05}.

\begin{figure} [!ht]
  \begin{center}
  \makeatletter
    \renewcommand{\p@figure}{figure\space}
  \makeatother
    \epsfig{file=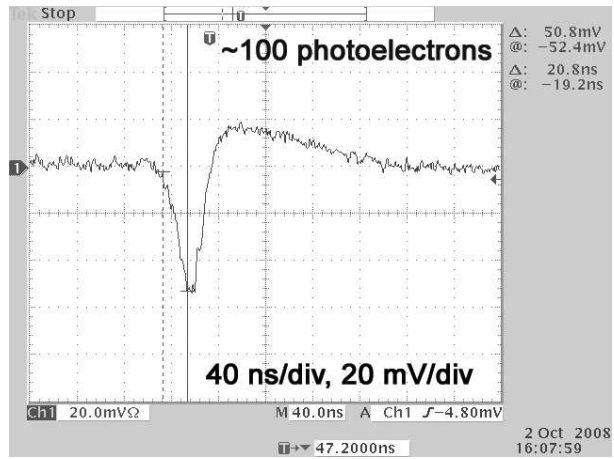, width=8cm}
    \caption{An example of a pulse recorded at the anode strips of MHSP of F-R-MHSP/GEM/MHSP multiplier.}
    \label{fig:FGM:mph:pulse}
  \end{center}
\end{figure}

An example of a single pulse induced by a photon-flash, recorded in a GPM comprising the F-R-MHSP/GEM/MHSP multiplier coupled to a CsI PC, is shown in \ref{fig:FGM:mph:pulse}. The pulses were recorded in the experimental setup described in \cite{lyashenko:07} with a fast current-sensitive preamplifier (0.5 ns rise-time) connected directly to the MHSP anode. The GPM's PC was illuminated by light flashes from a H$_2$-filled discharge lamp. The pulses were recorded at a multiplier's gain of $\sim3\cdot10^4$. The pulse in \ref{fig:FGM:mph:pulse} is of $\sim$100 photoelectrons; it has a characteristic rise-time of $\sim$20 ns).

\section{Summary}

The present paper summarized the various steps of the development of visible sensitive gas photomultipliers (VSGPMs), combining a thin-film (semitransparent) photocathode such as Cs$_3$Sb, K$_2$CsSb and Na$_2$KSb with a state-of-the-art cascaded electron multiplier: the F-R-MHSP/GEM/MHSP. The latter yielded a record avalanche-ion backflow level of $\sim$30 ions on the average per single-electron avalanche of 10$^5$ charges. Combining the recent knowledge gained on the ion-induced secondary effects probability \cite{lyashenko:09} with the ion-blocking capability of the novel multiplier \cite{lyashenko:07}, we could predict its operation stability at high gains, also in continuous mode. This was validated recently with such VSGPM operated at 700 Torr Ar/CH$_4$  (95/5), in an unsealed mode as reported in this work. At gains exceeding 10$^5$, fast photoelectron pulses, with $\sim$20 ns rise-time were recorded; the detector's stability was proved; the photocathode's QE remained stable over a month-period. Note that the VSGPM's characterization was carried out within the large-volume detection chamber of the UHV system; this somewhat limited the length of the experiments, due to the chemical degrading of the photocathodes, caused by (mostly surface) impurities.

The low electron-emission threshold of alkali-antimonide photocathodes makes them susceptible to ion feedback effects, imposing severe limits on the maximum reachable gains, as demonstrated in \cite{balcerzyk:03, moermann:thes}. The probability for ion-induced secondary electron emission was recently measured and calculated to be between 0.02 and 0.03 for Cs$_3$Sb, K$_2$CsSb and Na$_2$KSb PCs \cite{lyashenko:09}. This means 2-3 secondary electrons emitted for every 100 ions impinging on the PC, resulting in ion feedback after-pulses. To have feedback-free operation at gain 10$^5$  (to ensure single electron sensitivity) implies limiting the ion back flow fraction (IBF) to no more than $3\cdot10^{-4}$ \cite{lyashenko:09}. Cascaded GEM multipliers do not comply to this requirement on IBF; their IBF is at best 10 times higher \cite{bondar:03, sauli:06}. Some possibility to further limit the IBF in cascaded GEMs involves loss of single-electron collection efficiency, or, alternatively an operation in pulsed trigger mode \cite{mormann:03}. The requirement on IBF was recently shown to be fulfilled, with full single-electron collection efficiency and in continuous mode, in cascaded hole-multipliers incorporating GEM and MHSP elements. It was demonstrated \cite{lyashenko:07} that the required IBF value of $3\cdot10^{-4}$ was reached with a three-element F-R-MHSP/GEM/MHSP cascaded multiplier operated in atmospheric-pressure Ar/CH$_4$  (95/5), at total gain of $\sim$ 10$^5$  \cite{lyashenko:07}.

We described in details the deposition methods of Cs$_3$Sb, K$_2$CsSb and Na$_2$KSb photocathodes, and reported on additional studies such as photoemission into gas media, operation stability etc. In total we have produced 6 Na$_2$KSb PCs with peak QE of 15 to 25\%, 7 Cs$_3$Sb PCs with peak QE of 10 to 40\% and 28 K$_2$CsSb PCs. While the average QE-value of the latter was 28\%, most of the latest produced, after mastering the technology, had larger QE values - at the higher range of commercial photomultipliers in the spectral range of 313 - 546 nm.

As in the case of CsI PCs \cite{escada:07}, the electron emission from bi-alkali photocathodes into gas showed a significant dependence of the backscattering (and the resulting extraction efficiency $\varepsilon_{extr}$) on the wavelength, namely on the photoelectron energy. The extraction efficiency in Ar/CH$_4$  (95/5), at an electric field of 500V/cm, rose linearly from $\sim$ 40\% at 313 nm to $\sim$ 80\% (relative to the vacuum value) at a wavelength of 546 nm.

The stability of K$_2$CsSb and Cs$_3$Sb PCs was tested in somewhat unfavorable conditions (big volume chamber), yet, they showed stable QE over extended period of time ($\sim$ 4 weeks). The entire VSGPM was operated continuously in pulsed-light mode at a gain of 10$^5$  and a photon rate of $\sim$ 12 kHz/mm$^2$; it was followed for 14 hours (accumulated charge $\sim$ 125 $\mu$C) not showing any significant deterioration in performance. It yielded stable spark-less operation.

Fast, multiple-photon pulses recorded in a GPM with the F-R-MHSP/GEM/MHSP multiplier coupled to a CsI photocathode had typical rise-times of $\sim$ 20 ns. Based on previous studies with similar pulses in cascaded-GEM GPMs, we expect few-ns time resolution with single photons and sub-ns resolutions with multiple-photon light pulses \cite{chechik:08}.

The present work paves the way for further development, of the production of sealed VSGPMs, with flat geometry and area in the 100-1000 cm$^2$  ranges. They are expected to have several virtues: Fast photon imaging with single-photon sensitivity, large areas, operation under intense photon flux and at high magnetic fields. VSGPMs have the potential of competing with vacuum photon detectors in many fields of applications, such as in imaging Cherenkov light in particle physics and astrophysics, in  large-area neutrino detectors, in recording scintillation information in particle physics and medical imaging, etc. they might serve in advanced image intensifiers, etc. Though having lower quantum yields due to the backscattering effect in gas, VSGPMs may compete with solid-state photon detectors in size and cost, in single-photon sensitivity and in noise characteristics.

The progress from the small-area laboratory VSGPM prototype developed within this study, to large-area sealed devices is a technological challenge. While the production of large-area photocathodes and large-detector sealing has been mastered since long by industry, the production of large-area MHSP electrodes is yet to be demonstrated; progress in the production of large-area Micropattern gaseous detectors has been reached lately within the CERN-RD51 collaboration \cite{oliveira:09}.  Efforts should be undertaken to develop such or similar patterned hole-multipliers made of inert insulating materials; among possible candidates are glass, ceramic and silicon \cite{breskin:03}. Low-radioactivity substrate materials could be used in experiments aiming at the detection of rare events; among polymers, a possible candidate would be Cirlex \cite{gai:09}, investigated for thick-GEM (THGEM \cite{breskin:09}) multipliers.

The VSGPM's multiplier's performance in terms of ion blocking could possibly be further improved, replacing the last MHSP stage by a MICROMEGAS element \cite{giomataris:96}. The latter would ease the GPM's signal recording for imaging purposes. The MICROMEGAS multiplier provides a decent multiplication of about 10$^4$ and is known to have good ion backflow blocking \cite{colas:04}; production methods for large-area multipliers have been developed \cite{oliveira:09}.

An optimization of the gas filling could also reduce secondary effects, e.g. the addition of low ionization potential admixtures like Ethyl Ferrocene (EF) or TMAE could considerably reduce ion feedback. It was shown \cite{biteman:01} that in a GPM filled with He/CH$_4$  comprising a Cs$_3$Sb PC and a capillary-plate multiplier, an addition of EF vapor to the gas mixture improved the maximal achievable gain of the device.

\acknowledgments
We would like to thank Dr. D. M\"{o}rmann for the development, within his Ph.D. thesis work, of the experimental system and some of the methods described in this article; we are thankful to M. Klin for his technical support. This work was partly supported by the Israel Science Foundation, grant No 402/05, by the MINERVA Foundation, grant 8566 and by Project POCTI/FP/63441/2005 through FEDER and FCT (Lisbon). A. Breskin is the W.P. Reuther Professor of Research in The Peaceful Use of Atomic Energy.

\bibliographystyle{JHEP}
\bibliography{publications}

\providecommand{\href}[2]{#2}\begingroup\raggedright\begin{thebibliography}{10}

\bibitem{chechik:08}
R.~Chechik and A.~Breskin, {\it Advances in gaseous photomultipliers},  {\em
  Nucl. Instr. and Meth. A} {\bf 595} (2008) 116.

\bibitem{edmends:88}
J.~Edmends {\em et.~al.}, {\it Investigation of prototype gas-filled
  photomultipliers},  {\em Nucl. Instr. and Meth. A} {\bf 273} (1988) 145.

\bibitem{peskov:99}
V.~Peskov {\em et.~al.}, {\it First attempts to combine capillary tubes with
  photocathodes},  {\em Nucl. Instr. and Meth. A} {\bf 433} (1999) 492.

\bibitem{shefer:02}
E.~Shefer {\em et.~al.}, {\it Photoelectron transport in csi and csbr coating
  films of alkali antimonide and csi photocathodes},  {\em J. Appl. Phys.} {\bf
  92} (2002) 4758.

\bibitem{balcerzyk:03}
M.~Balcerzyk {\em et.~al.}, {\it Methods of preparation and performance of
  sealed gas photomultipliers for visible light},  {\em IEEE Trans. Nucl. Sci.}
  {\bf 50} (2003) 847.

\bibitem{sauli:97}
F.~Sauli, {\it {GEM}: {A} new concept for electron amplification in gas
  detectors},  {\em Nucl. Instr. and Meth. A} {\bf 386} (1997) 531.

\bibitem{veloso:00}
J.~F. C.~A. Veloso {\em et.~al.}, {\it Micromegas, a multipurpose gaseous
  detector},  {\em Rev. Sci. Inst. A} {\bf 71} (2000) 2371.

\bibitem{mormann:03}
D.~M\"{o}rmann {\em et.~al.}, {\it {GEM}-based gaseous photomultipliers for
  {UV} and visible photon imaging},  {\em Nucl. Instr. and Meth. A} {\bf 504}
  (2003) 93.

\bibitem{moermann:thes}
D.~M\"{o}rmann, {\it Study of novel gaseous photomultipliers for {UV} and
  visible light},  {\em {P}h.{D}. thesis, {W}eizmann {I}nstitute of {S}cience,
  {I}srael}.

\bibitem{bondar:03}
A.~Bondar {\em et.~al.}, {\it Study of ion feedback in multi-{GEM} structures},
   {\em Nucl. Instr. and Meth. A} {\bf 496} (2003) 325.

\bibitem{sauli:06}
F.~Sauli {\em et.~al.}, {\it Ion feedback suppression in time projection
  chambers},  {\em Nucl. Instr. and Meth. A} {\bf 560} (2006) 269.

\bibitem{maia:04}
J.~M.~F. Maia {\em et.~al.}, {\it Avalanche-ion back-flow reduction in gaseous
  electron multipliers based on {GEM}/{MHSP}},  {\em Nucl. Instr. and Meth. A}
  {\bf 523} (2004) 334.

\bibitem{lyashenko:06}
A.~Lyashenko {\em et.~al.}, {\it Avalanche-ion back-flow reduction in gaseous
  electron multipliers based on {GEM}/{MHSP}},  {\em JINST} {\bf 1} (2006)
  P10004.

\bibitem{lyashenko:07}
A.~Lyashenko {\em et.~al.}, {\it Further progress in ion back-flow reduction
  with patterned gaseous hole-multipliers},  {\em JINST} {\bf 2} (2007) P08004
  and references therein.

\bibitem{lyashenko:09}
A.~Lyashenko {\em et.~al.}, {\it Ion-induced secondary electron emission from
  {K}-{C}s-{S}b, {N}a-{K}-sb and {C}s-{S}b photocathodes and its relevance to
  the operation of gaseous avalanche photomultipliers},  {\em J. Appl. Phys.}
  (2009) in preparation.

\bibitem{shefer:98}
E.~Shefer {\em et.~al.}, {\it Laboratory production of efficient
  alkali-antimonide photocathodes},  {\em Nucl. Instr. and Meth. A} {\bf 411}
  (1998) 383.

\bibitem{braem:03}
A.~Braem {\em et.~al.}, {\it Aging of large-area {C}s{I} photocathodes for the
  {ALICE} {HMPID} prototypes},  {\em Nucl. Instr. and Meth. A} {\bf 515} (2003)
  307.

\bibitem{engstrom:80}
R.~Engstrom, {\em Photomultiplier handbook}.
\newblock RCA Corp., 1980.

\bibitem{escada:07}
J.~Escada {\em et.~al.}, {\it A {M}onte {C}arlo study of backscattering effects
  in the photoelectron emission from {C}s{I} into {CH}$_{4}$ and
  {A}r-{CH}$_{4}$ mixtures},  {\em JINST} {\bf 2} (2007) P08001.

\bibitem{buzulutskov:00}
A.~Buzulutskov {\em et.~al.}, {\it The {GEM} photomultiplier operated with
  noble gas mixtures},  {\em Nucl. Instr. and Meth. A} {\bf 443} (2000) 164.

\bibitem{dimauro:96}
A.~D. Mauro {\em et.~al.}, {\it Photoelectron backscattering effects in
  photoemission from {C}s{I} into gas media},  {\em Nucl. Instr. and Meth. A}
  {\bf 371} (1996) 137.

\bibitem{rodionov:02}
I.~Rodionov {\em et.~al.}, {\it Hybrid gaseous photomultipliers},  {\em Nucl.
  Instr. and Meth. A} {\bf 478} (2002) 384.

\bibitem{ahn:01}
S.~Ahn {\em et.~al.}, {\it {GEM}-type detectors using {LIGA} and etchable glass
  technologies},  {\em Lawrence Berkeley National Laboratory} {\bf Paper
  LBNL-47782} (2001).

\bibitem{azmoun:06}
B.~Azmoun {\em et.~al.}, {\it A study of gain stability and charging effects in
  {GEM} foils},  {\em {IEEE} Nuclear Science Symposium Conference Record} {\bf
  6} (2006) 3847.

\bibitem{breskin:05}
A.~Breskin {\em et.~al.}, {\it Ion-induced effects in {GEM} and {GEM}/{MHSP}
  gaseous photomultipliers for the {UV} and the visible spectral range},  {\em
  Nucl. Instr. and Meth. A} {\bf 553} (2005) 46 and references therein.

\bibitem{oliveira:09}
R.~de~Oliveira, {\it Recent achievements and projects in {L}arge {MPGD}s
  \href{http://indico.cern.ch/getFile.py/access?contribId=0\&resId=1\&material%
Id=slides\&confId=47406}{http://indico.cern.ch/getFile.py/access?contribId=0\&%
resId=1\&materialId=slides\&confId=47406}},  {\em Talk given at the 2nd {RD}-51
  {C}ollaboration {M}eeting}.

\bibitem{breskin:03}
A.~Breskin {\em et.~al.}, {\it Recent advances in gaseous imaging
  photomultipliers},  {\em Nucl. Instr. and Meth. A} {\bf 513} (2003) 250.

\bibitem{gai:09}
M.~Gai {\em et.~al.}, {\it Toward {A}pplication of a {T}hick {G}as {E}lectron
  {M}ultiplier ({THGEM}) {R}eadout for a {D}ark {M}atter {D}etector},
  \href{http://xxx.lanl.gov/abs/0706.1106}{{\tt arXiv:0706.1106}}.

\bibitem{breskin:09}
A.~Breskin {\em et.~al.}, {\it A concise review on {THGEM} detectors},  {\em
  Nucl. Instr. and Meth. A} {\bf 598} (2009) 107.

\bibitem{giomataris:96}
Y.~Giomataris {\em et.~al.}, {\it {MICROMEGAS}: a high-granularity
  position-sensitive gaseous detector for high particle-flux environments},
  {\em Nucl. Instr. and Meth. A} {\bf 376} (1996) 29.

\bibitem{colas:04}
P.~Colas {\em et.~al.}, {\it Ion backflow in the {MICROMEGAS} {TPC} for the
  future linear collider},  {\em Nucl. Instr. and Meth. A} {\bf 535} (2004)
  226.

\bibitem{biteman:01}
V.~Biteman {\em et.~al.}, {\it Position sensitive gaseous photomultipliers},
  {\em Nucl. Instr. and Meth. A} {\bf 471} (2001) 205.

\end{thebibliography}\endgroup
\end{document}